\documentclass[12pt,twoside]{article}
\usepackage{correl}
\usepackage[punctsep]{collref}
\def\TheTitle{Kondo physics and the Mott transition}
\def\TheHeading{Kondo\&Mott}
\def\TheAuthor{Michele Fabrizio}
\def\TheInstitute{International School for Advanced Studies, SISSA}
\def\TheAddress{Via Bonomea 265, I-34136 Trieste, Italy}
\def\TheChapter{5}                       

\usepackage{subfigure}

\newcommand{\be}{\begin{equation}}
\newcommand{\ee}{\end{equation}}
\newcommand{\bea}{\begin{eqnarray}}
\newcommand{\eea}{\end{eqnarray}}
\newcommand{\ba}{\begin{eqnarray*}}
\newcommand{\ea}{\end{eqnarray*}}
\newcommand{\dagga}{{\phantom{\dagger}}}

\newcommand{\bq}{\mathbf{q}}

\newcommand{\bk}{\mathbf{k}}

\newcommand{\bp}{\mathbf{p}}

\newcommand{\dis}{\displaystyle}

\newcommand{\up}{\uparrow}
\newcommand{\down}{\downarrow}
\newcommand{\fract}[2]{\frac{\dis \;#1\;}{\dis \;#2\;}}

\newcommand{\eqn}[1]{(\ref{#1})}

\newcommand{\ep}{{\epsilon}}

\newcommand{\bw}{\begin{widetext}}
\newcommand{\ew}{\end{widetext}}

\begin{document}
\MakeTitle           
\section{A brief recall of Landau-Fermi-liquid theory}
Landau's Fermi-liquid theory~\cite{Landau-1,Landau-2} 
explains why interacting fermions, despite repelling each other by  Coulomb interaction, almost always display thermodynamic and transport properties similar to those of non-interacting particles, which is e.g. the reason of success of the Drude-Sommerfeld description of normal metals in terms of free-electrons. \\
The microscopic justification of Landau's Fermi-liquid theory, see e.g. Ref.~ \cite{NozieresPR1962-1,NozieresPR1962-2}, is a beautiful and elegant realisation of what we would now denote as a renormalizable field theory. I will not go through all details of such theory, but just emphasise few aspects linked to the main subject of the present notes.  \\
The step zero of Landau's Fermi-liquid theory is the assumption\footnote{This assumption can be actually verified order by order in perturbation theory, which however does not guarantee that the perturbation series is convergent} that the fully interacting single-particle Green's function close to Fermi, $|k-k_F|\ll k_F$ and $|\ep|\ll\ep_F$, 
includes a coherent and an incoherent component, namely
\be
G(i\ep,\bk) \simeq G_\text{coh.}(i\ep,\bk)+G_\text{incoh.}(i\ep,\bk) = \fract{Z_\bk}{i\ep -\ep_\bk} + G_\text{incoh.}(i\ep,\bk)\,,\label{Green}
\ee
where $\ep$ are Matsubara frequencies, $\ep_\bk$ is measured with respect to the Fermi energy $\ep_F$,  
and $Z_\bk\leq 1$ is the so-called quasi-particle residue. The Green's function continued in the complex frequency plane $i\ep\to z\in\mathbb{C}$ has therefore the simple  pole singularity of $G_\text{coh.}(z,\bk)$ plus, generically, a branch cut on the real axis brought by 
$G_\text{incoh.}(z,\bk)$. By definition $G(z=\ep+i0^+,\bk)-G(z=\ep-i0^+,\bk) = -2\pi i\,\mathcal{N}(\ep,\bk)$, where $\mathcal{N}(\ep,\bk)$ is the single-particle density of states, which therefore reads, according to Eq.~\eqn{Green},
\be
\mathcal{N}(\ep,\bk) = Z_\bk\,\delta\big(\ep-\ep_\bk\big) + \mathcal{N}_\text{incoh.}(\ep,\bk)\,.\label{A}
\ee 
Since $\mathcal{N}(\ep,\bk)$ has unit integral over $\ep$, the incoherent component has weight $1-Z_\bk$. 
The meaning of Eq.~\eqn{A} is that an electron added to the system transforms, with weight $Z_\bk$, into a \textit{quasi-particle} excitation that propagates with dispersion $\ep_\bk$, but also into a bunch of other excitations that do not propagate coherently. The distinction between coherent and incoherent becomes sharper analysing the analytic behaviour, in the mathematical sense of a distribution, of the product \index{Landau-Fermi liquid theory}
\be
R(i\ep,\bk;i\omega,\bq) \equiv G(i\ep+i\omega,\bk+\bq)\,G(i\ep,\bk)\,,\label{R}
\ee
that enters the calculation of low-temperature linear response functions in the low-frequency, $i\omega=\omega+i0^+$ with 
$\omega\ll \ep_F$, long-wavelength, $|\bq|\ll k_F$,  limit, the measurable quantities which are the ultimate goal of the theory.  Indeed, one finds by elementary calculations that 
\be
\begin{split}
R(i\ep,\bk;i\omega,\bq) &=  G_\text{coh.}(i\ep+i\omega,\bk+\bq)\,G_\text{coh.}(i\ep,\bk) 
+ R_\text{incoh.}(i\ep,\bk;i\omega,\bq)\\
&\simeq - \fract{\partial f\left(\epsilon_\bk\right)}{\partial \epsilon_\bk}\;
\delta\left(i\epsilon\right)\, Z_\bk^2\, 
\fract{\epsilon_{\bk+\bq}-\epsilon_\bk}{i\omega - \epsilon_{\bk+\bq} + \epsilon_\bk} 
+ R_\text{incoh.}(i\ep,\bk;i\omega,\bq)\\
&\equiv R_\text{coh.}(i\ep,\bk;i\omega,\bq) + R_\text{incoh.}(i\ep,\bk;i\omega,\bq)\,.
\end{split}
\ee
The crucial point that distinguishes $R_\text{coh.}(i\ep,\bk;i\omega,\bq)$ from 
$R_\text{incoh.}(i\ep,\bk;i\omega,\bq)$ is that the former is evidently non-analytic in the origin, $\omega=\bq=0$, 
while the latter is assumed to be analytic. In other words, while the limiting value of 
$R_\text{incoh.}(i\ep,\bk;i\omega,\bq)$ at $\omega=\bq=0$ is unique, 
\be
\lim_{\omega\to 0}\,\lim_{\bq\to 0} \,R_\text{incoh.}(i\ep,\bk;i\omega,\bq) =
\lim_{\bq\to 0} \,\lim_{\omega\to 0}\,R_\text{incoh.}(i\ep,\bk;i\omega,\bq) \equiv 
R_\text{incoh.}(i\ep,\bk)\,,
\ee 
that of 
$R_\text{coh.}(i\ep,\bk;i\omega,\bq)$ is instead not unique and depends how the limit is taken: 
\be
\begin{split}
\lim_{\omega\to 0}\,\lim_{\bq\to 0} \,R_\text{coh.}(i\ep,\bk;i\omega,\bq) 
&\equiv R_\text{coh.}^\omega(i\ep,\bk) = 0\,,\\
\lim_{\bq\to 0} \,\lim_{\omega\to 0}\,R_\text{coh.}(i\ep,\bk;i\omega,\bq) 
&\equiv R_\text{coh.}^q(i\ep,\bk) = \fract{\partial f\left(\epsilon_\bk\right)}{\partial \epsilon_\bk}\;
\delta\left(i\epsilon\right)\, Z_\bk^2\,,
\end{split}
\ee
where the two different limits are conventionally indicated by the superscripts $\omega$ and $q$.  It thus 
follows that 
\be
\begin{split}
R^\omega(i\ep,\bk) &= R_\text{incoh.}(i\ep,\bk)\,,\\
R^q(i\ep,\bk) &= R_\text{coh.}^q(i\ep,\bk) + R_\text{incoh.}(i\ep,\bk)\,.
\end{split}
\ee
\begin{figure}[t!]
 \centering
 \includegraphics[width=0.5\textwidth]{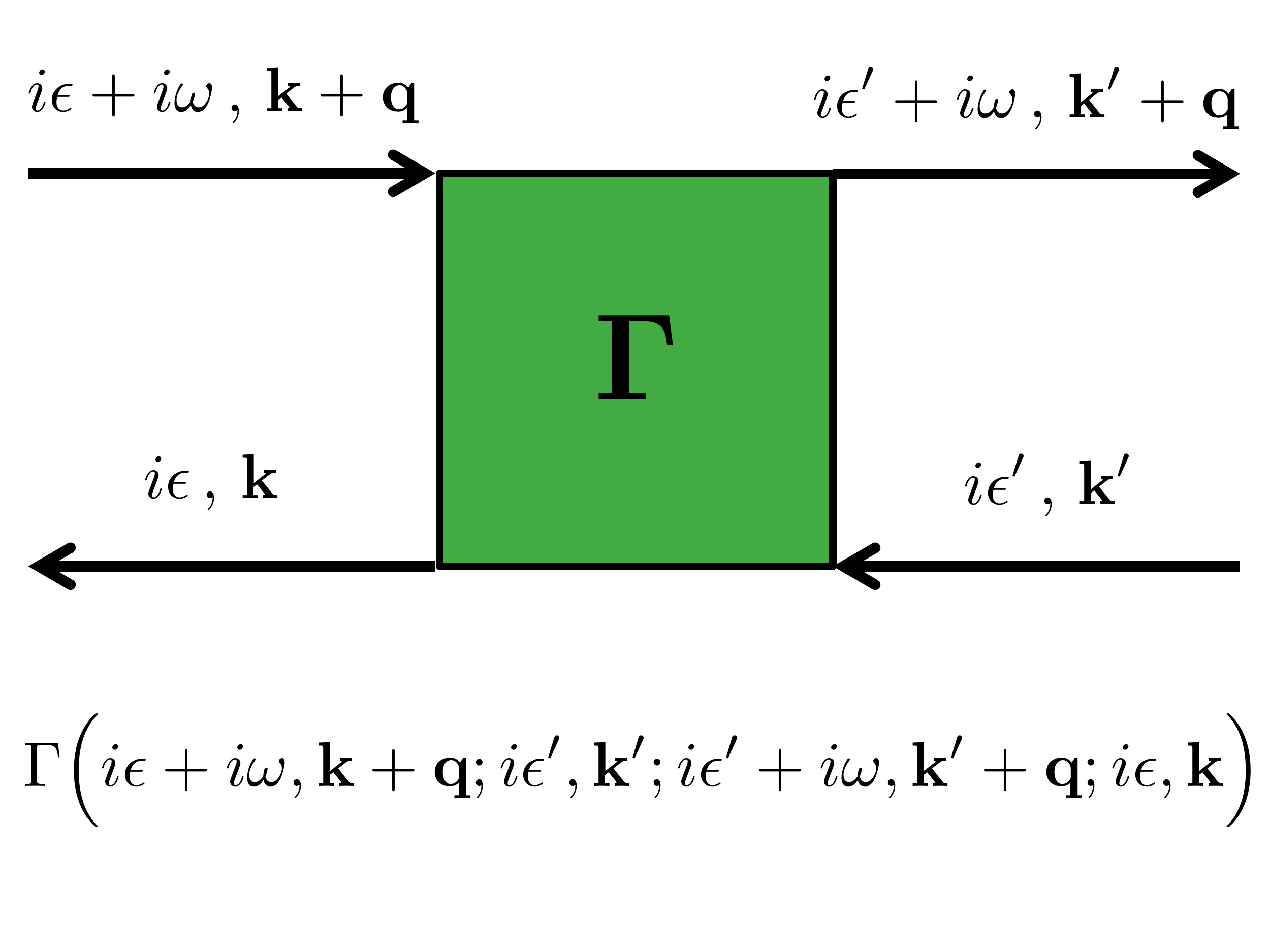}\hfill
\includegraphics[width=0.4\textwidth]{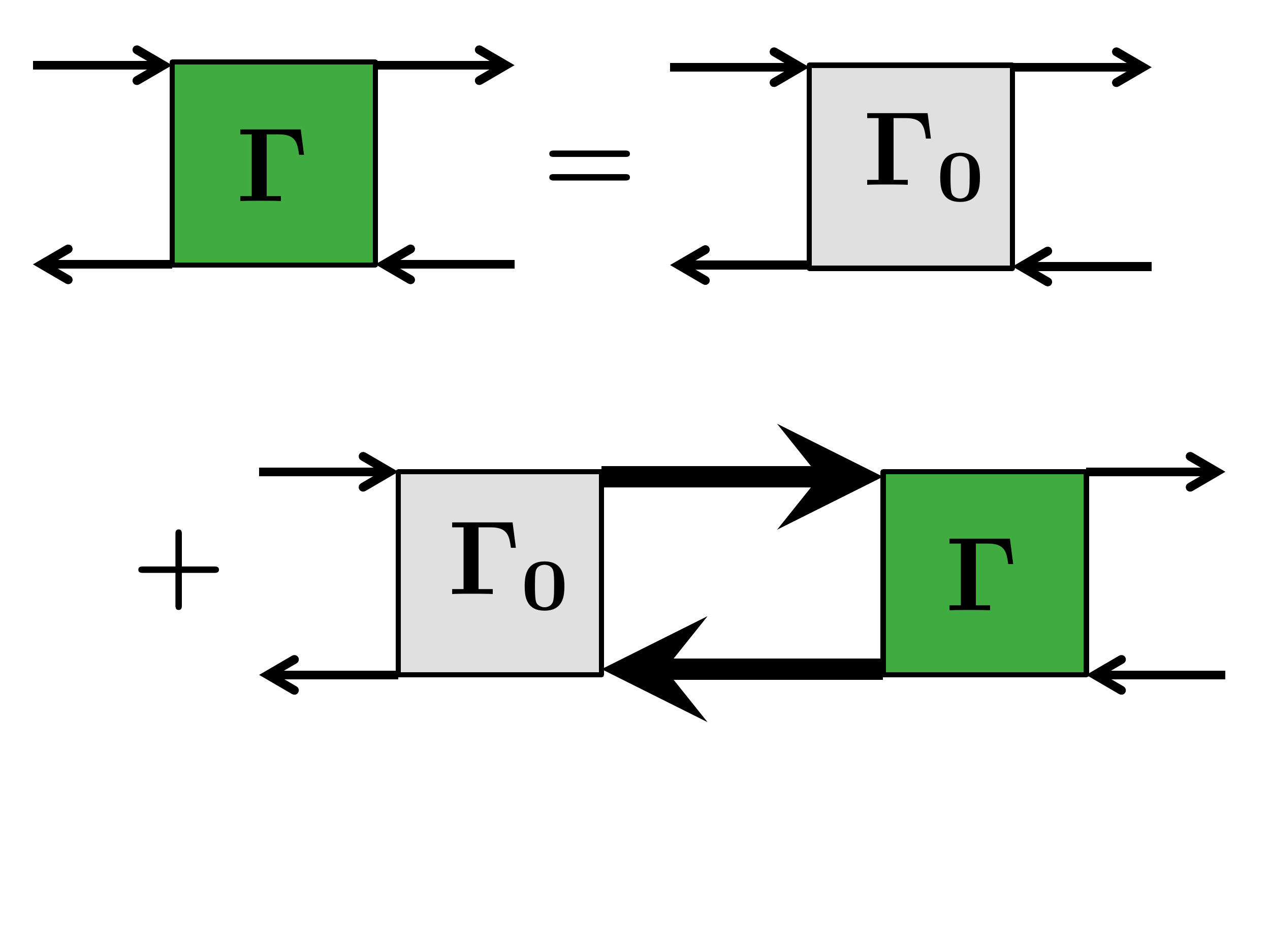}  
 \caption{Left panel: diagrammatic representation of the interaction vertex $\Gamma$ in the particle-hole channel with frequency and momentum transferred $\omega$ and $\bq$, respectively.
 Right panel: diagrammatic representation of the Bethe-Salpeter equation. $\Gamma_0$ is the irreducible vertex, and the product of the two internal Green's functions, the two tick lines, is by definition the distribution $R$.}
 \label{Gamma}
\end{figure}
The next important step within Landau's Fermi-liquid theory is to absorb the completely unknown 
$R_\text{incoh.}(i\ep,\bk)$ into few parameters. I will not repeat throughly what is well explained in many other places, but just sketch how it works in the case of the Bethe-Salpeter equation that relates the reducible vertex in the particle-hole channel $\Gamma$ to the irreducible one $\Gamma_0$ and to $R$, see Fig.~\ref{Gamma}. 
To simplify the notations, I will not explicitly indicate external and internal variables, frequencies and momenta, and indicate by $\odot$ the summation over the internal ones. With those conventions the Bethe-Salpeter equation reads
\be
\Gamma = \Gamma_0 + \Gamma_0\odot R\odot \Gamma\,,
\ee
so that $\Gamma^\omega = \Gamma_0 + \Gamma_0\odot R^\omega\odot \Gamma^\omega$ and 
$\Gamma^q = \Gamma_0 + \Gamma_0\odot R^q\odot \Gamma^q$ with the same $\Gamma_0$, since 
by construction $\Gamma_0$ is analytic at $\omega=\bq=0$. Solving for $\Gamma_0$ one readily finds that 
\be
\begin{split}
\Gamma &= \Gamma^\omega + \Gamma^\omega\odot \Big(R-R^\omega\Big)\odot\Gamma = 
\Gamma^q + \Gamma^q\odot \Big(R-R^q\Big)\odot\Gamma\,,
\end{split}
\ee
where 
\be
\begin{split}
R-R^\omega &\simeq - \fract{\partial f\left(\epsilon_\bk\right)}{\partial \epsilon_\bk}\;
\delta\left(i\epsilon\right)\, Z_\bk^2\, 
\fract{\epsilon_{\bk+\bq}-\epsilon_\bk}{i\omega - \epsilon_{\bk+\bq} + \epsilon_\bk} \;,\\
R-R^q &\simeq - \fract{\partial f\left(\epsilon_\bk\right)}{\partial \epsilon_\bk}\;
\delta\left(i\epsilon\right)\, Z_\bk^2\, 
\fract{i\omega}{i\omega - \epsilon_{\bk+\bq} + \epsilon_\bk} \;,
\end{split}\label{2-R}
\ee
do not involve anymore $R_\text{incoh.}$,  at the expenses of introducing two unknown objects, 
$\Gamma^\omega$ and $\Gamma^q$. Those are actually not independent since, e.g., 
\be
\begin{split}
\Gamma^q &= \Gamma^\omega + \Gamma^\omega\odot \Big(R^q-R^\omega\Big)\odot\Gamma^q \,.
\end{split}
\ee
Conventionally one uses $\Gamma^\omega$ and define the Landau's $f$-parameters through
\footnote{Note that the two expressions in Eq.~\eqn{2-R} are finite only at $\ep=0$, so that one only needs the vertex at zero Matsubara frequencies in the calculation of linear response functions.} 
\be
f_{\bk\bp} = Z_\bk\,Z_\bp\, \Gamma^\omega(0,\bk;0,\bp;0,\bp;0,\bk)\,,
\ee
where I introduced back the external variables according to the figure~\ref{Gamma}. Exploiting Ward's identities one can derive the known Fermi liquid expressions, in terms of the above-defined $f$-parameters 
and of the unknown dispersion $\ep_\bk$, of the linear response functions at small $\omega$ and $\bq$ for all conserved quantities, for which we refer e.g. to Ref.~ \cite{ NozieresPR1962-1,NozieresPR1962-2}. 

\section{Ordinary Kondo physics at the Mott transition}
\label{Ordinary Kondo physics at the Mott transition}
\begin{figure}[t!]
 \centering
 \includegraphics[width=0.6\textwidth]{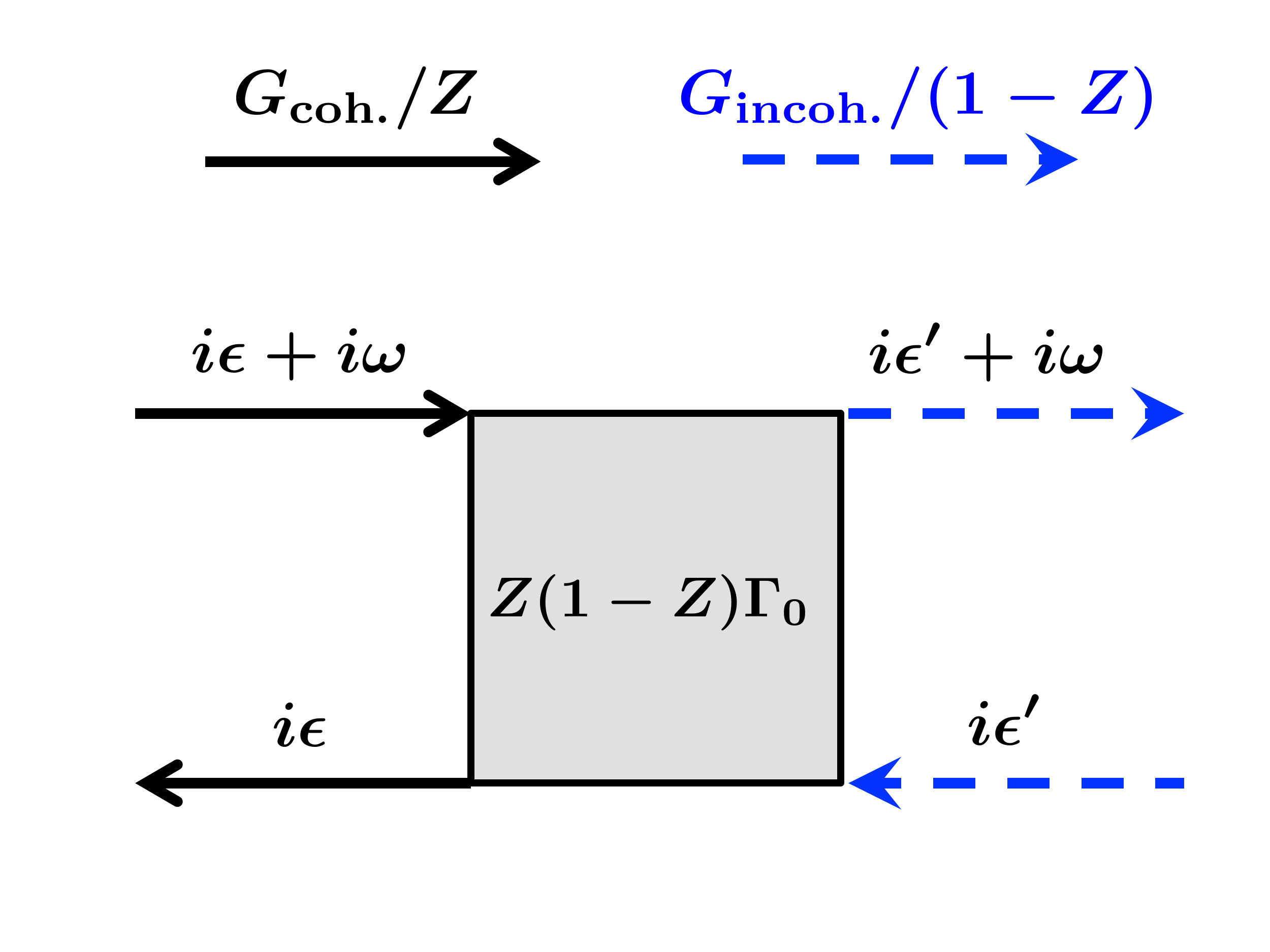}
 \caption{Coherent and incoherent components of the Green's function, and the irreducible scattering vertex among them that can transfer low frequency $\omega$.} 
 \label{Green-fig}
 \end{figure}
Commonly one cannot model the incoherent background and therefore the best one can do is invoking Landau's Fermi-liquid theory to get rid of $R_\text{incoh.}$ and $G_\text{incoh.}$. There is however a situation where we can proceed a bit further. Let us imagine to be in a strongly correlated metal phase close to a Mott transition, i.e. a metal-to-insulator transition driven by the electron-electron repulsion. In this circumstance we can grasp what the incoherent background represents. Indeed, 
in the Mott insulating phase the low-energy coherent component of the Green's 
function, $G_\text{coh.}$, has disappeared, while the incoherent $G_\text{incoh.}$ must describe the atomic-like excitations of the insulator. I shall assume that the insulator has low-energy degrees of freedom, which cannot involve the charge, since its fluctuations are suppressed, but may involve the spin and/or, if present, the orbital degrees of freedom. We can thus imagine that, in the metal phase contiguous to the Mott insulator, $G_\text{incoh.}$ still describes the same atomic-like excitations, though coexisting with low-energy quasiparticle excitations. I shall indicate $G_\text{coh.}(i\ep,\ep_\bk)/Z_\bk 
=1/\big(i\ep-\ep_\bk\big)$ and $G_\text{incoh.}/(1-Z)$ with solid and dashed lines, respectively, see Fig.~\ref{Green-fig}.  Accordingly, the irreducible vertex becomes $Z(1-Z)\Gamma_0$.
Among all irreducible scattering processes that couple among each other coherent and incoherent components, the only one that can transfer low energy is that depicted in Fig.~\ref{Green-fig}. Since charge fluctuations cost energy in the insulator, that scattering vertex $Z(1-Z)\Gamma_0$ acts only in the spin and/or orbital channels. For instance, in the single-band case that vertex should describe a spin exchange between itinerant quasiparticles and localised moments. In other words, the strongly correlated metal close to the Mott transition should behave similarly to a Kondo lattice model, i.e. conduction electrons coupled by a spin exchange $J=Z(1-Z)\Gamma_0$ to local moments, with the major difference that $J$ is not an Hamiltonian parameter but it is self-consistently determined by the fully interacting theory. \index{Landau-Fermi liquid theory!emergence of Kondo physics}\\

The above very crude arguments that suggest a similarity between the physics of the Kondo effect and that of the Mott transition turn into a rigorous proof in lattices with infinite coordination number, $z\to\infty$, limit in which  the so-called dynamical mean-field theory (DMFT)~\cite{DMFT} becomes exact. Within DMFT a lattice model is mapped onto an Anderson impurity model coupled to a bath. The mapping is exact for $z\to\infty$ provided a self-consistency condition between the local Green's function of the bath and the 
impurity Greens function is fulfilled.\index{dynamical mean field theory} Even though the mapping strictly holds only for $z\to\infty$, the previous heuristic arguments point to a more general validity, with the due differences coming from the fact that spatial fluctuations, which can be neglected in infinitely coordinated lattices, grow in importance as the coordination number decreases. \\
The obvious step further is therefore how to export the well-established knowledge of the Kondo effect to the physics of the Mott transition. Here one has to face two problems: 
\begin{itemize}
\item[1.] Even when the two models are rigorously mappable onto each other, i.e. in the limit of infinite coordination number,  yet the mapping holds only under a self-consistency condition. How does such a condition affect the physics?
\item[2.] When the lattice has finite coordination, spatial correlations cannot be neglected anymore, e.g. the single-particle self-energy acquires momentum dependence. How does the physics across the Mott transition change? 
\end{itemize}
In what follows I will just touch the first issue, which is also the simpler, assuming a model in an infinitely coordinated lattice. 
\subsection{Role of the DMFT self-consistency condition}
\begin{figure}[t!]
 \centering
 \includegraphics[width=0.6\textwidth]{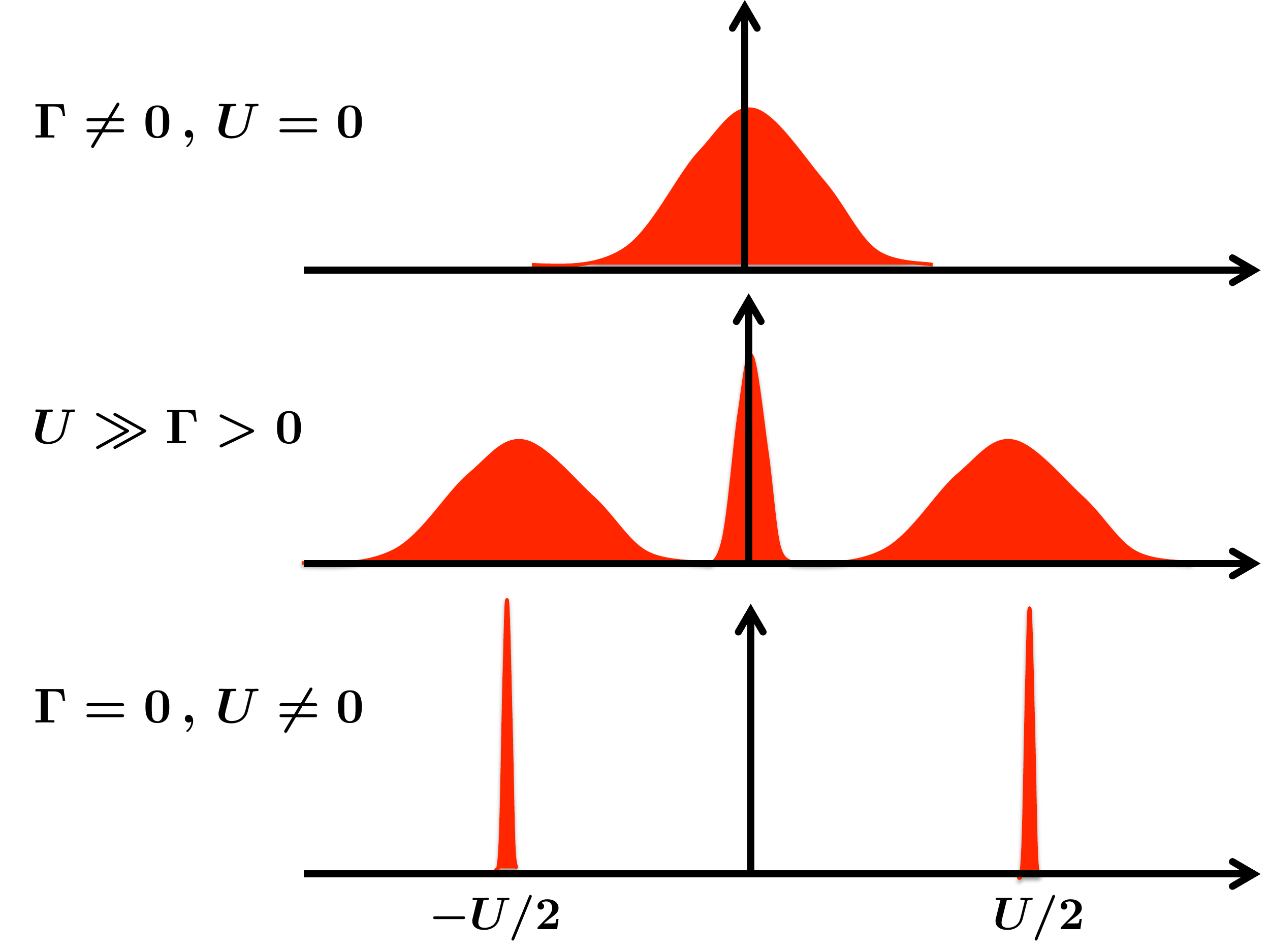}
 \caption{Sketch of the impurity density of states in three limiting cases.} 
 \label{DOS}
 \end{figure}
Let us start from the simplest case of the single-band Hubbard model at half-filling. Here the mapping is simply that onto a single-orbital Anderson impurity model (AIM) with Hamiltonian
\be
\begin{split}
\mathcal{H}_\text{AIM} &= \sum_{\bk\sigma}\,\ep_\bk\,c^\dagger_{\bk\sigma}c^\dagga_{\bk\sigma} 
+ \sum_{\bk\sigma}\,V_\bk\,\Big(c^\dagger_{\bk\sigma}d^\dagga_\sigma + 
d^\dagger_\sigma c^\dagga_{\bk\sigma}\Big) + \fract{U}{2}\big(n-1\big)^2\,,
\end{split}\label{AIM}
\ee
where $c^\dagga_{\bk\sigma}$ and $d^\dagga_\sigma$ are the annihilation operators of 
the conduction and impurity electrons, respectively,  the bath dispersion $\ep_\bk$ is measured with respect to the chemical potential and finally 
$n=\sum_\sigma\, d^\dagger_\sigma  d^\dagga_\sigma$ is the occupation of the impurity level.  The model \eqn{AIM} depends actually on two quantities: the Hubbard repulsion $U$ and the so-called hybridisation function 
\be
\Gamma(\ep) = \pi\sum_\bk\, \left|V_\bk\right|^2\,\delta\big(\ep-\ep_\bk)\,.\label{Gamma(ep)}
\ee
When $U=0$ and $\Gamma(\ep)\not=0$, the impurity density of states (DOS), $\mathcal{N}(\ep)$, which was a $\delta$-function centred at the chemical potential $\ep=0$ in the absence of hybridisation with the bath, becomes in its presence a Lorentzian of width 
$\Gamma \simeq \Gamma(0)$, see top panel in Fig.~\ref{DOS}. On the contrary, when $U\not = 0$ and $\Gamma(\ep)=0$, the isolated impurity is singly occupied in its ground state, so that its DOS, which measures at zero temperature the probability of removing, at $\ep<0$, or adding, at $\ep>0$, an impurity electron, displays two $\delta$-peaks at $\ep=\pm U/2$, see bottom panel in Fig.~\ref{DOS}, where $U/2$ is the energy cost of the empty or doubly occupied  impurity states. Those side peaks are known as the Hubbard bands. When both $U$ and $\Gamma(\ep)$ are non zero, the DOS actually displays both features, namely a roughly Lorentzian peak at $\ep=0$, whose width is renormalised downwards by $U$, $\Gamma\to\Gamma_* = Z\,\Gamma$ with $Z<1$, and two side-peaks centred at 
$\ep=\pm U/2$ that are broadened by hybridisation by an amount $\propto \Gamma$, see middle panel in Fig.~\ref{DOS}.\\
Remarkably, the central peak exists for any value of $U$, even if bigger than any other energy scale. If $U$ 
is very large, the impurity is singly occupied by either a spin up or down electron and thus essentially behaves as a spin-1/2 local moment. Nonetheless, the system can still gain hybridisation energy by screening the impurity spin through the conduction electrons, what is named as the \textit{Kondo effect}.\index{Kondo effect} As a result, a tiny fraction of the impurity DOS is promoted at the chemical potential $\ep=0$ and gives rise to a very narrow peak , the so-called Kondo or Abrikosov-Suhl resonance. Its width $\Gamma_* = Z\,\Gamma\ll\Gamma$ defines the so-called Kondo temperature $T_K=\Gamma_*$, above which screening is not anymore effective. In other words, for temperatures $T>T_K$ the impurity behaves effectively as a \textit{free} spin-1/2 and the Kondo resonance has disappeared.
\index{Kondo effect!single impurity}\\
This is in brief the physical behaviour of the single-orbital AIM without any DMFT self-consistency. The latter roughly amounts to requiring that the hybridisation function, $\Gamma(\ep)$ of Eq.~\eqn{Gamma(ep)}, has a similar shape to the impurity DOS. Therefore, once self-consistency is imposed, the effective impurity model is defined by a $\Gamma(\ep)$ that also displays a peak of width 
$\Gamma_*$ at Fermi separated from two higher-energy Hubbard side-bands, see middle panel in Fig.~\ref{DOS}. As $U$ increases the peak at Fermi of $\Gamma(\ep)$ thus becomes narrower and narrower 
until, at a critical $U_c$, Kondo screening of the impurity spin by the conduction bath is not anymore sustainable and the Abrikosov-Suhl resonance disappears, i.e. $\Gamma_*\to 0$. Above $U_c$ the impurity DOS, which is also the local Green's function of the lattice model, only displays two well separated Hubbard bands; the system is therefore turned into a Mott insulator. \index{dynamical mean field theory!role of self-consistency}\\

\textit{-- The first important role of the self-consistency is thus to push down at finite $U=U_c$ what in the impurity model without self-consistency happens only at $U=\infty$, i.e. the disappearance of the Kondo resonance. }\\

In the single-orbital AIM the impurity magnetic susceptibility, 
$\chi_\text{imp.}\sim 1/\Gamma_*$,
grows more and more as $\Gamma_*\to 0$. This suggests that the lattice model counterpart should develop some kind of magnetic instability before the Mott transition. Such instability is forbidden in the Anderson impurity model without DMFT self-consistency, since spin $SU(2)$ cannot be locally broken, but it might occur when self-consistency is enforced because a global spontaneous $SU(2)$ symmetry breaking is instead allowed. This is not at all unexpected. Indeed, local moments develop as the metal moves close to the Mott transition; these moments must order one way or another to get free of their $\ln 2$ entropy.  \\

\textit{-- We can therefore argue that another important 
effect of self-consistency is to transform the impurity instabilities into genuine bulk instabilities of the corresponding lattice model, which may thus drive a transition into symmetry broken phases prior or concurrently with the Mott transition. }

\section{Exotic Kondo physics at the Mott transition}

The last conjecture entails appealing scenarios which might be realised in lattice models that map within DMFT onto impurity models with a richer phase diagram than the simple single-orbital one. There is indeed a whole zoo of impurity models with varied physical properties. 
I note that the metal phase close to the Mott transition corresponds by DMFT self-consistency 
to an Anderson impurity model with $U\gg \Gamma_*$ that suppresses valence fluctuations. In this regime the model becomes equivalent to a Kondo model where the impurity effectively behaves as a local moment with spin magnitude $S$, generically greater that 1/2, and eventually endowed with additional internal degrees of freedom brought, e.g., by orbital degeneracy as in the case of partially filled $d$ or $f$ shells.\\
Kondo models describing a spin-$S$ impurity, with no other internal degrees of freedom, coupled to $k$-channels of spin-1/2 conduction electrons are divided into: 
(1) $k>2S$ overscreened Kondo models; (2) $k=2S$ screened Kondo models; and (3) $k<2S$ underscreened Kondo models. Even though overscreened Kondo models are potentially interesting since they display instabilities in several channels~\cite{Affleck-NPB1991}, yet they will never appear in DMFT since by construction a lattice model in infinitely coordinated lattices maps unavoidably onto an impurity model in which the number of degrees of freedom of the impurity is the same as that of the conduction bath, i.e. $k=2S$ in the above example.\\
\begin{figure}[t!]
 \centering
 \includegraphics[width=0.5\textwidth]{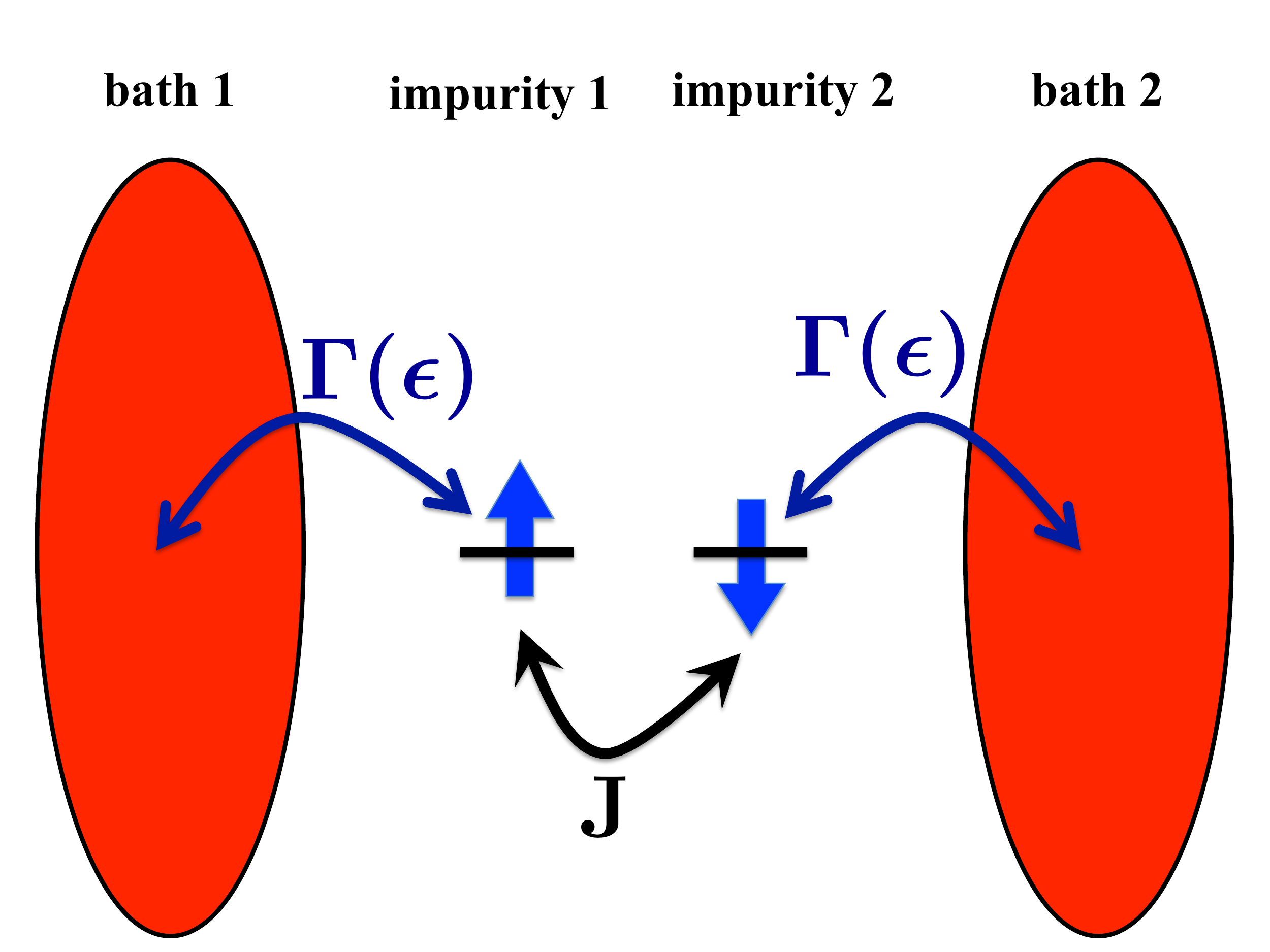}
 \caption{The two impurity Anderson model. Each impurity is coupled by hybridisation, parametrised here by the hybridisation function $\Gamma(\ep)$, with its own conduction bath. In addition the two impurities are coupled among each other by an antiferromagnetic spin-exchange.}
 \label{2AIM}
\end{figure}

\subsection{The two impurity model}
There is however another much more promising class of impurity models characterised by the existence of internal degrees of freedom of the impurity besides the spin , and, more importantly, by an additional Hamiltonian parameter $J$ that is able to quench those degrees of freedom and thus competes against Kondo screening. Out of this competition a rich phase diagram emerges, which generally includes a quantum critical point or a narrow crossover region that separate the phase in which the impurity degrees of freedom are quenched by Kondo screening from that in which quenching is due to $J$.\\
The best-known representative of this class is the two-impurity Anderson model~\cite{Jones&Varma1987,Jones&Varma1988,Jones&Varma1989}, which I shall now discuss as the prototypical example. This model is depicted in Fig.~\ref{2AIM}; it consists of two equivalent single-orbital AIM's in which the two impurities are not only hybridised each to its own bath, but also coupled among each other by an antiferromagnetic spin-exchange $J$. The Hamiltonian reads
\be
\mathcal{H}_\text{2AIM} = \mathcal{H}_\text{AIM-1}+\mathcal{H}_\text{AIM-2} + J\,\mathbf{S}_1\cdot\mathbf{S}_2\,,\label{H2AIM}
\ee
where $ \mathcal{H}_\text{AIM-a}$ is the AIM Hamiltonian Eq.~\eqn{AIM} of the impurity 
$a=1,2$, and 
\[
\mathbf{S}_a = \fract{1}{2}\,\sum_{\alpha\beta}\,d^\dagger_{a\alpha}\,\boldsymbol{\sigma}\,
d^\dagga_{b\beta}\,,
\]
its spin operator with $\boldsymbol{\sigma}=\big(\sigma_x,\sigma_y,\sigma_z\big)$ the Pauli matrices. 
The Hamiltonian \eqn{H2AIM} has three relevant parameters, $U$, the hybridisation function 
$\Gamma(\ep)$, by definition equal for each impurity, and the exchange $J$. If $J=0$, each impurity is Kondo screened by its bath on the energy scale given by the Kondo temperature $T_K$. If, on the contrary, 
$\Gamma(\ep)=0$ but $J\not=0$, the impurities are decoupled from the baths but coupled among each other into a spin-singlet configuration. Both cases are stable in the sense that no degeneracy is left to be lifted. 
If all parameters are finite, the Kondo screening, with scale $T_K$, competes against the direct exchange $J$. Therefore, if $T_K\gg J$, the system prefers to Kondo screen each impurity with its bath. On the contrary, 
if $J\gg T_K$, the two impurities lock into a singlet state that is transparent to the conduction electrons. These two limiting cases, which I shall denote as screened, $T_K\gg J$, and unscreened, $J\gg T_K$, 
 correspond to two different phases separated by a genuine quantum critical point (QCP) at $T_K\sim J$. Its critical properties have been uncovered in great details~\cite{L&A-PRL1992,L&A&J-PRB1995}.
\index{Kondo effect!two impurities} Specifically, at the QCP the model display logarithmically singular susceptibilities in several channels:
\begin{itemize}
\item[(1)] the "antiferromagnetic" channel defined by the operators 
\be
\boldsymbol{\Delta}_\text{AFM} = \mathbf{S}_1-\mathbf{S}_2\,,\label{1}
\ee
\item[(2)] the "hybridisation" channels 
\be
\Delta_x = \sum_\sigma\, \Big(d^\dagger_{1\sigma}d^\dagga_{2\sigma}+H.c.\Big)\,,\qquad 
\Delta_y = i\,\sum_\sigma\, \Big(d^\dagger_{1\sigma}d^\dagga_{2\sigma}-H.c.\Big)\,,\label{2}
\ee
\item[(3)] the spin-singlet Cooper channel
\be
\Delta = d^\dagger_{1\up}d^\dagger_{2\down}+ d^\dagger_{2\up}d^\dagger_{1\down}\,,
\label{3}
\ee
and its hermitian conjugate $\Delta^\dagger$. 
\end{itemize}
\begin{figure}[t!]
 \centering
 \includegraphics[width=0.5\textwidth]{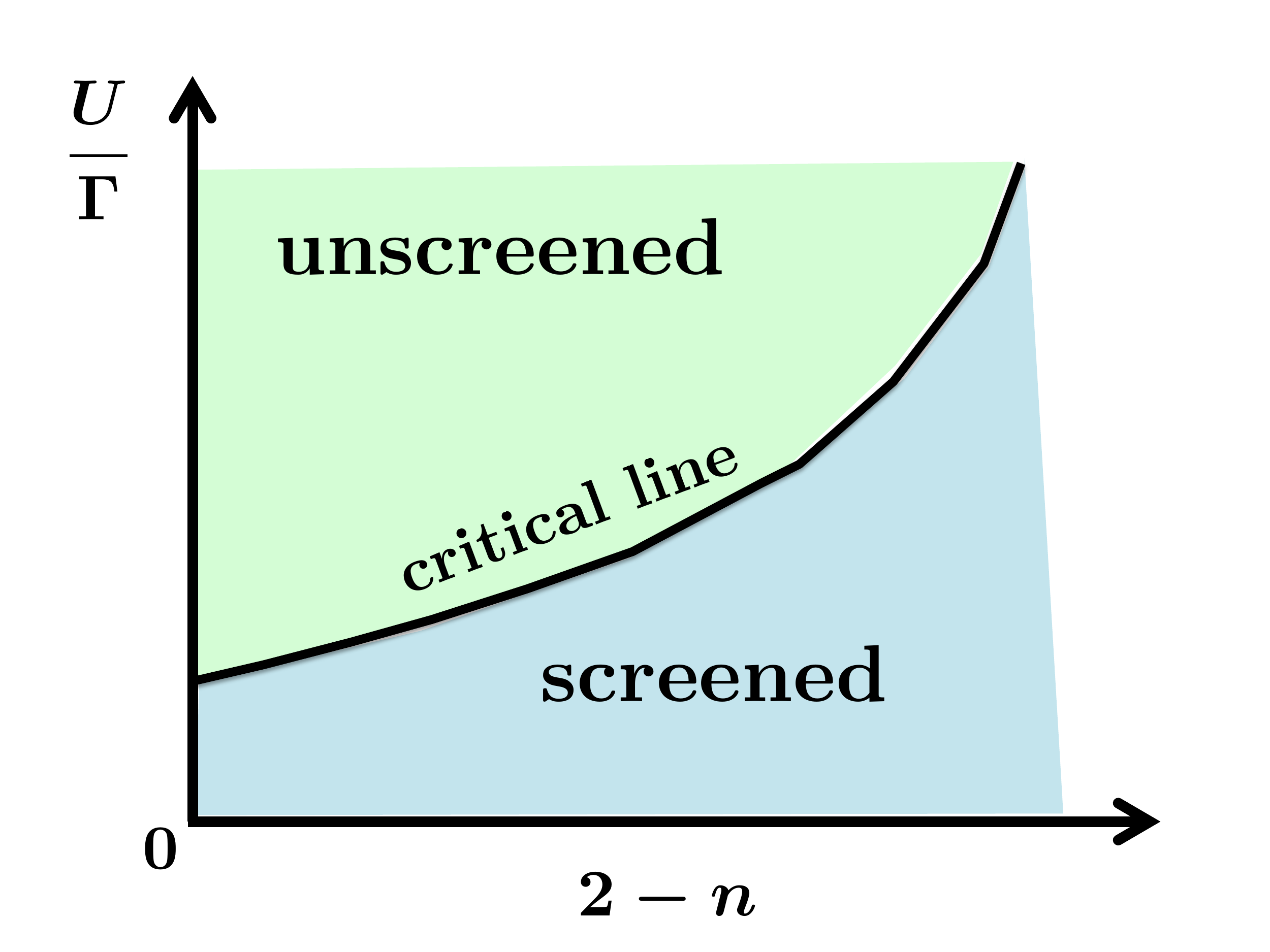}
 \caption{Sketch of phase diagram at fixed $J/U$ as function of $U/\Gamma$ and 
 doping $2-n$ of the two impurities away from half-filling. }
 \label{phd}
\end{figure}
On the contrary, the impurity charge susceptibility is not singular since charge fluctuations are suppressed by the large $U$. As a consequence, the 
QCP is stable upon \textit{doping} the impurity site, which corresponds to changing the position of the impurity level so that 
$n\equiv \langle n_1+n_2\rangle \not= 2$. The phase diagram is schematically shown in Fig.~\ref{phd}. One can observe that the QCP at half-filling, $n=2$, is actually the endpoint of a whole critical line that moves upwards in $U/\Gamma$ at fixed $J/U$ away from half-filling. 
In other words, if one starts from the unscreened phase at half-filling and dopes the impurity, at some doping the critical line will be crossed. \index{Kondo effect!two impurities!phase diagram}\\

The dynamical behaviour of the impurity DOS across the QCP has been uncovered quite in detail~\cite{LorenzoPRB2004,Nutshell}. The vicinity of the QCP is controlled by two energy scales. One is smooth across the transition: it was denoted as $T_+$ in Ref.~\cite{LorenzoPRB2004} and was found to be of the order  $\text{max}\big(T_K,J\big)$, where $T_K$ is the Kondo temperature at $J=0$. The other energy scale $T_-$ measures the deviation from 
the QCP.  I recall that the Kondo temperature $T_K$ at $J=0$ is defined by $\Gamma\equiv\Gamma(0)$ 
and $U$ according to 
\[
T_K(\Gamma,U)  = U \sqrt{\fract{\Gamma}{2U}\,}\,\exp\bigg(-\fract{\pi U}{8\Gamma}-\fract{\pi \Gamma}{2U}
\bigg)\,,
\]
and decreases by decreasing $\Gamma$ or increasing $U$. Let us for instance assume that $J$ and $U$ are fixed while $\Gamma$ varies. In this case the QCP is identified by 
$\Gamma=\Gamma_c$ such that $T_K(\Gamma_c,U)\simeq J$, and $T_-\propto \big(\Gamma - \Gamma_c\big)^2$, vanishing quadratically at the transition. 
\begin{figure}[t!]
 \centering
 \includegraphics[width=0.7\textwidth]{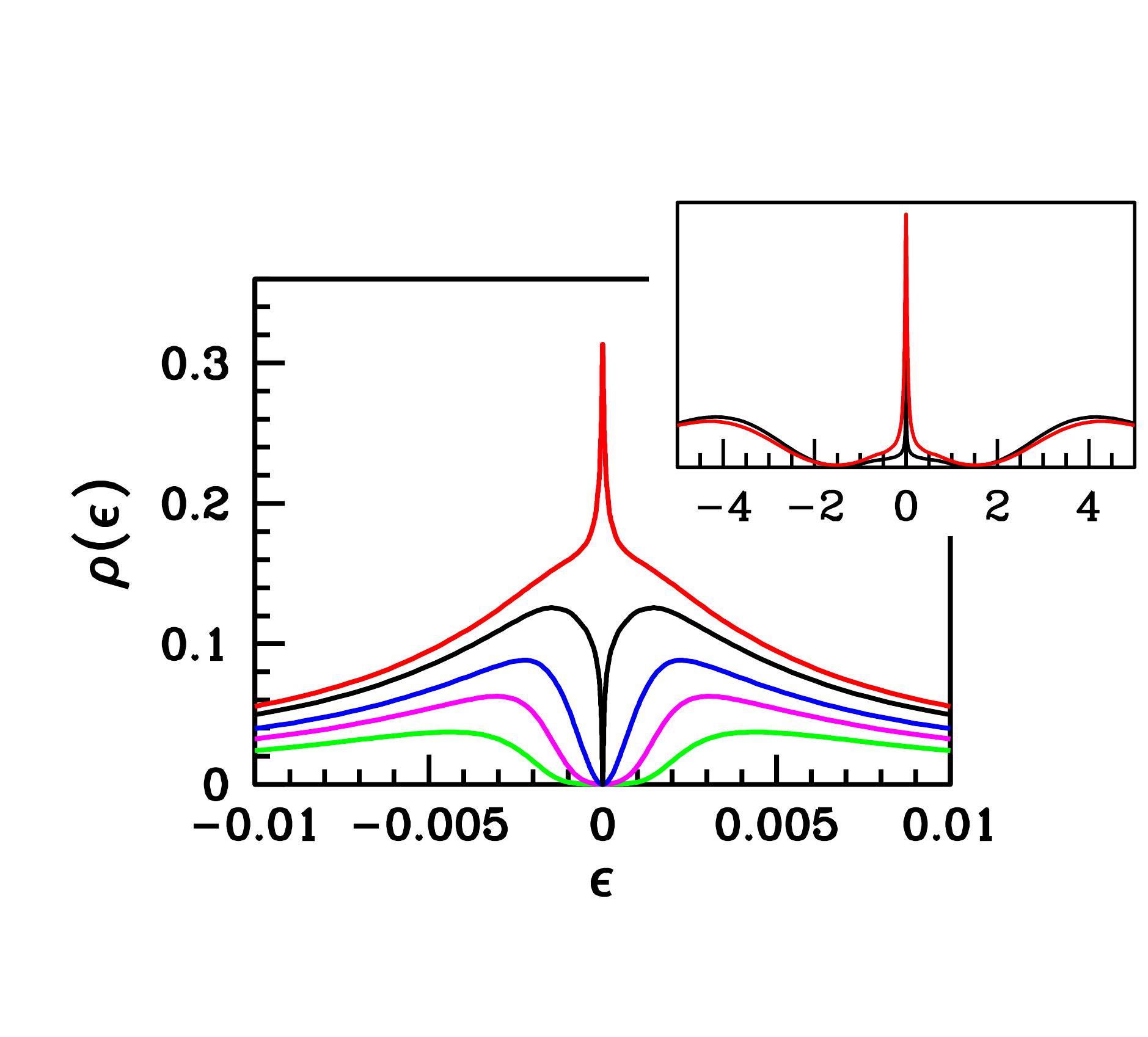}
 \vspace{-1cm}
 \caption{Low-energy DOS of the two impurity Anderson model across the phase transition. The calculation are done at $U = 8$, $J = 0.00125$ and the curves, from top to bottom,  corresponds to $\Gamma= 0.44, 0.42, 0.4, 0.35, 0.3$ in units 
the conduction bandwidth. The critical point is at $\Gamma=\Gamma_c\simeq 0.43956$. 
 In the screened phase a narrow Kondo resonance is present, red curve. In the unscreened phase instead the DOS has a pseudo gap. In the inset the DOS on a larger scale is shown, where the Hubbard bands are visible 
 [From Ref.~\cite{Nutshell}]. The colours of the curves correspond to those in the main panel. }
 \label{2AIMDOS}
\end{figure}
It was found~\cite{LorenzoPRB2004} that the impurity DOS as obtained by numerical renormalisation group 
is well fit at low energy by the expression\index{Kondo effect!two impurities!spectral properties}  
\be
\mathcal{N}_\pm(\ep) = \fract{1}{2\pi\Gamma}\bigg(
\fract{T_+^2}{\ep^2+T_+^2} \pm \fract{T_-^2}{\ep^2+T_-^2}\bigg)\,,
\label{N(ep)}
\ee
where the $+$ refers to the Kondo screened phase, $\Gamma>\Gamma_c$, and the $-$ to the unscreened one, $\Gamma<\Gamma_c$. Right at the QCP
\be
\mathcal{N}_c(\ep) = \fract{1}{2\pi\Gamma_c}\;
\fract{T_+^2}{\ep^2+T_+^2} \;. \label{N*}
\ee
I note that the DOS in the screened phase is the sum of two Lorentzian's, one of width $T_+$ and a much narrower one of width $T_-$ that vanishes at the QCP. Here only the broader peak remains. On the unscreened side of the transition, the DOS is the difference of two Lorentzian's, and its value at the chemical potential vanishes -- a pseudo gap emerges by the disappearance of Kondo screening. I further note that such a pseudo gap is not to be confused with the much larger one that separates lower from upper Hubbard bands, see the inset of Fig.~\ref{2AIMDOS}. The two are indeed controlled by different energy scales, $U$ the latter and $T_-\ll U$ the former.\\ 
When the impurity is doped, i.e. its occupation number $n$ deviates from half-filling $n=2$, the low-energy DOS was found~\cite{LorenzoPRB2004} to be still of the form Eq.~\eqn{N(ep)}, 
\be
\mathcal{N}_\pm(\ep) = \fract{\cos^2\nu}{2\pi\Gamma}\Bigg(
\fract{T_+^2+\mu_\pm^2}{\big(\ep+\mu_\pm\big)^2+T_+^2} \pm 
\cos 2\nu\;\fract{T_-^2}{\ep^2+T_-^2}\Bigg)\,,
\label{N(ep)-mu}
\ee
where $\mu_\pm=\pm T_+\,\sin 2\nu$ measures the deviation away from half-filling. Remarkably, the narrower Lorentzian remains peaked at the chemical potential, so that the pseudo gap in the unscreened phase is pinned at Fermi.  

\begin{figure}[t!]
 \centering
 \includegraphics[width=0.5\textwidth]{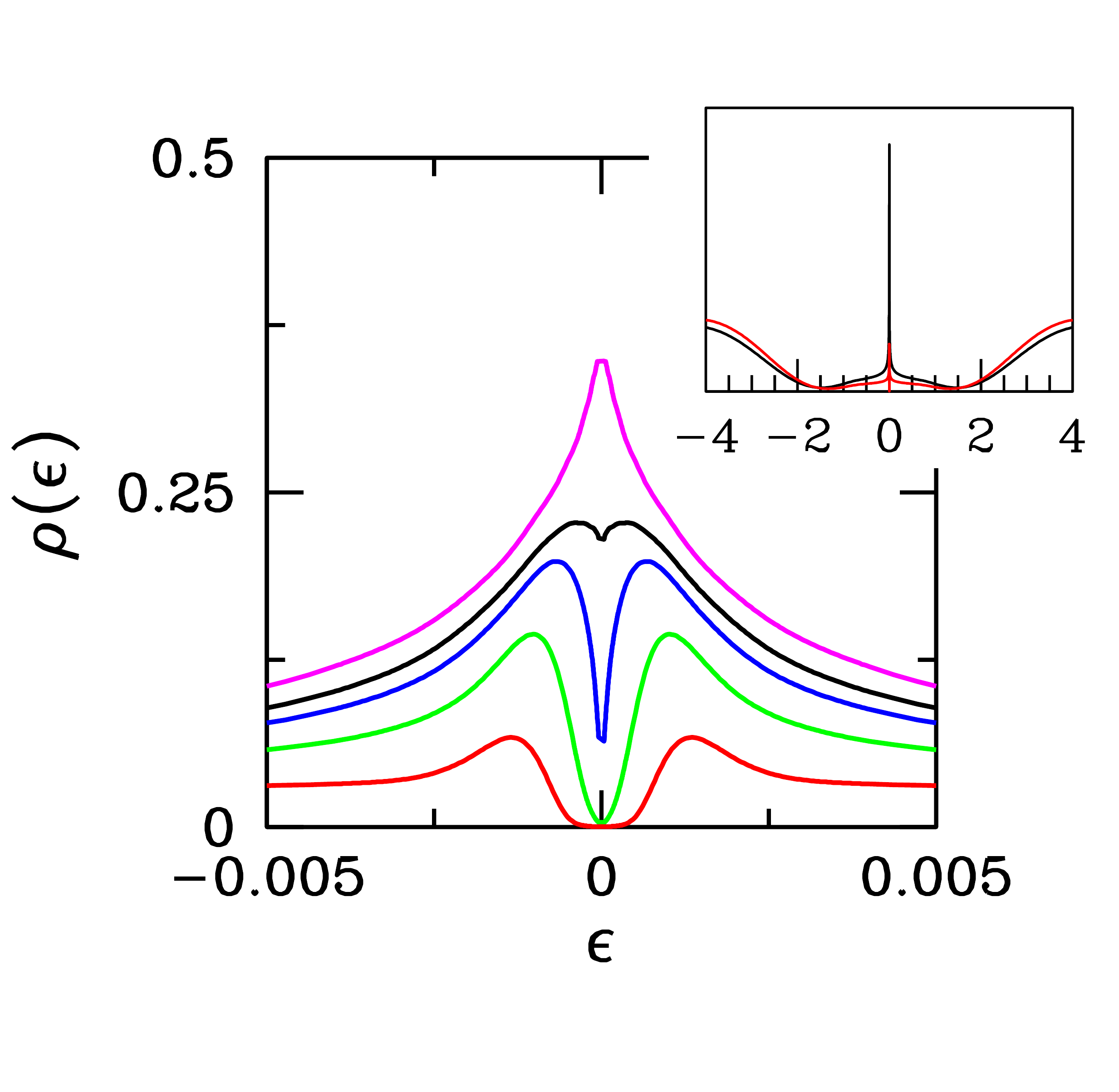}
 \vspace{-1cm}
 \caption{Low-energy DOS of the two impurity Anderson model 
 Eq.~\eqn{H2AIM-1} for $U=8$, $t_\perp=0.05$, and, 
 from top to bottom, $\Gamma= 0.5, 0.47, 0.45, 0.4, 0.3$ in units 
the conduction bandwidth. In the inset the DOS on a larger scale is shown, where the Hubbard bands are visible 
 [From Ref.~\cite{Nutshell}]. The colours of the curves correspond to those in the main panel. }
 \label{2AIMDOSbis}
\end{figure}

\subsubsection{Explicit symmetry breaking} 
Let us now discuss more in detail the role of the operators in Eq.~\eqn{2}, focusing in particular on $\Delta_x$ that describes a direct real hybridisation among the impurities. This operator breaks the $U(1)$ orbital symmetry 
\be
\begin{split}
d^\dagga_{1\sigma}&\to \text{e}^{i\phi}\;d^\dagga_{1\sigma}\,,
\qquad c^\dagga_{1\bk\sigma} \to \text{e}^{i\phi}\;c^\dagga_{1\bk\sigma}
\,,\\
d^\dagga_{2\sigma}&\to \text{e}^{-i\phi}\;d^\dagga_{2\sigma}\,,
\qquad c^\dagga_{2\bk\sigma} \to \text{e}^{-i\phi}\;c^\dagga_{2\bk\sigma}
\,,
\end{split}\label{U(1)}
\ee
of the original Hamiltonian \eqn{H2AIM}, where $d^\dagga_{a\sigma}$ 
and $c^\dagga_{a\bk\sigma}$ are the annihilation operators of the $a=1,2$ impurity and conduction electrons, respectively. In the language of critical phenomena, such $U(1)$ symmetry breaking is therefore a relevant perturbation that spoils the QCP. In other words, if the Hamiltonian were not  invariant under that symmetry, there would not be anymore a quantum phase transition but just a crossover between the screened and unscreened phases. \\

Suppose we consider the following Hamiltonian instead of that in Eq.~\eqn{H2AIM}
\be
\mathcal{H}'_\text{2AIM} = \mathcal{H}_\text{AIM-1}+\mathcal{H}_\text{AIM-2} - 
t_\perp\,\Delta_x
\,.\label{H2AIM-1}
\ee
This Hamiltonian is not invariant under the $U(1)$ symmetry in \eqn{U(1)}, therefore should not possess the above QCP. On the other hand, if $U\gg t_\perp$, $\mathcal{H}'_\text{2AIM}$ of Eq.~\eqn{H2AIM-1} 
can be mapped onto $\mathcal{H}_\text{2AIM}$ with $J=4t_\perp^2/U$, which instead has the QCP. 
How can we reconcile this apparent paradox? The answer is quite instructive. Indeed, $\mathcal{H}'_\text{2AIM}$ of Eq.~\eqn{H2AIM-1} does map onto $\mathcal{H}_\text{2AIM}$ 
of Eq.~\eqn{H2AIM} with $J=4t_\perp^2/U$, but just at leading order in $1/U$.  What really prevents the system from encountering the QCP are symmetry variant sub-leading terms, with coupling constant $h_x\propto \Gamma^2\,t_\perp/U^2\ll J$, which actually correspond to a direct hybridisation among the two conduction baths. In other words, a hierarchy of energy scales emerges naturally at large $U$ from the single $t_\perp$: $J$, which alone would drive the system across the phase transition, and a much smaller scale $h_x\ll J$ that allows the system crossing from the screened phase to the unscreened one without eventually 
passing through the QCP.  In the language of critical phenomena, we could state that although the system does not cross the QCP, yet it gets very close to it. Practically, this implies that the quantum phase transition turns into a very sharp crossover between screened and unscreened phases that, for many purposes, it is indistinguishable from a phase transition. In Fig.~\ref{2AIMDOSbis} the impurity DOS 
of the Hamiltonian \eqn{H2AIM-1} is shown, with parameters $U=8$ and $t_\perp=0.05$, in units of the conduction bandwidth, such that $4t_\perp^2/U$ is equal to the value of $J$ in Fig.~\ref{2AIMDOS}.
We first observe that in this case the DOS is always finite at 
$\ep=0$, though very small in the unscreened phase. In addition we can note that, despite the opening of the incomplete pseudo gap does not occurs through a phase transition, nonetheless it is extremely sharp.  

In conclusion, we can thus interpret the above results as those of the model Eq.~\eqn{H2AIM} at $J=4t_\perp^2/U$ in the presence of a small $h_x\ll J$ symmetry breaking field. 
From this perspective, inside the unscreened phase $J$ is responsible of the pseudo gap opening, while the much smaller $h_x$ of the partial filling of that same gap. 
A question immediately arises. How is it possible that the unscreened phase, even though pseudo-gapped, namely despite the impurities have a vanishing quasiparticle residue $Z=0$, 
is able to respond so efficiently to the small symmetry breaking field $h_x\ll J$? 

\subsubsection{How can a pseudo-gap symmetry invariant phase develop a symmetry variant order parameter?} 
This question has been addressed in Ref.~\cite{MarcoPRB2008} recognising a curious analogy between this problem and that of disordered $s$-wave superconductors. I will briefly sketch such relationship since I believe it reveals a basic   
feature that can be exported in many other contexts.\\
The expression Eq.~\eqn{N(ep)} of the low-energy DOS corresponds  
to the impurity Green's function in Matsubara frequencies
\be
\mathcal{G}(i\ep) = \fract{1}{2\Gamma}\,\bigg(
\fract{T_+}{i\ep +iT_+\,\text{sign}(\ep)}
\pm  \fract{T_-}{i\ep +iT_-\,\text{sign}(\ep)}\bigg)
+ \mathcal{G}_\text{incoh.}(i\ep)\,,\label{Gpm}
\,,
\ee
where $\pm$ refers, as before, to the screened and unscreened phases, respectively, and $\mathcal{G}_\text{incoh.}(i\ep)$ is the high-energy contribution from the Hubbard sidebands. In turns the Green's function satisfies the Dyson equation 
\be
\begin{split}
\mathcal{G}(i\ep)^{-1} = \mathcal{G}_0(i\ep)^{-1} -\Sigma(i\ep) 
= i\ep + i\Gamma\,\text{sign}(\ep) -\Sigma(i\ep)\,,
\end{split}
\ee
where $\mathcal{G}_0(i\ep)$ is the non-interacting Green's function and 
$\Sigma(i\ep)$ the impurity self-energy. 
We thus find that the impurity self-energy at low-energy 
and in the unscreened phase has the following expression
\be
\begin{split}
\Sigma(i\ep) &\simeq i\ep -\fract{i}{4\ep}\;
\fract{T_+ T_-}{T_+-T_-} - \fract{i}{4}\;
\fract{T_+ + T_-}{T_+-T_-}\;\text{sign}(\ep) 
-i\,\fract{\ep}{T_+-T_-}\\
&\equiv i\ep - i\,\fract{\ep}{Z(i\ep)}
\,,
\end{split}\label{Sigma-1}
\ee 
and diverges at $\ep\to 0$. The quasiparticle residue 
$Z(i\ep)$ thus vanishes at $\ep=0$.\\ 
Let us consider instead a disordered metal in the normal phase, whose 
self-energy is 
\be
\Sigma(i\ep) = \fract{i}{2\tau}\,\text{sign}(\ep) \equiv 
i\ep - i\ep\,\eta(i\ep)\,,
\ee
where $\tau$ is the relaxation time, and $\eta(i\ep)$ diverges at $\ep\to 0$, which, in analogy with Eq.~\eqn{Sigma-1}, can be interpreted as a vanishing quasiparticle residue. In the superconducting phase the self-energy acquires anomalous components and must be written as a two by two matrix
\be
\hat{\Sigma}(i\ep) = 
\begin{pmatrix}
\Sigma_{11}(i\ep) & \Sigma_{12}(i\ep)\\
\Sigma_{21}(i\ep) & \Sigma_{22}(i\ep) 
\end{pmatrix},\label{Sigma-2}
\ee  
where $\Sigma_{22}(i\ep)=-\Sigma_{11}(-i\ep)$ 
and $\Sigma_{21}(i\ep) = \Sigma_{12}(-i\ep)^*$. Because of the perfect cancellation of the disorder-induced corrections to the self-energy and to the vertex in the $s$-wave Cooper channel, superconductivity regularises the singularities brought by disorder below some low-energy scale $\Delta$, the superconducting gap,  leading to the following expressions of the self-energy matrix elements\index{Kondo effect!two impurities!pseudo-gap physics} 
\be
\begin{split}
\Sigma_{11}(i\ep) &= i\ep - i\ep\,\eta
\Big(i\,\sqrt{\ep^2+\Delta^2\,}\Big)\,,\\
\Sigma_{12}(i\ep) &= \Delta\,\eta
\Big(i\,\sqrt{\ep^2+\Delta^2\,}\Big)\,.
\end{split}\label{Sigma-3}
\ee
A famous consequence of Eq.~\eqn{Sigma-3}, known as \textit{Anderson theorem}, is that $T_c$ is independent of disorder strength, if weak, 
which readily follows from the BCS gap equation in the presence of an attraction $\lambda$
\be
1 = \lambda\,\fract{T}{V}\,\sum_{i\ep}\,\sum_\bk\,
\fract{\eta
\Big(i\,\sqrt{\ep^2+\Delta^2\,}\Big)}
{\big(\ep^2+\Delta^2\big)\,\eta
\Big(i\,\sqrt{\ep^2+\Delta^2\,}\Big)^2 +\ep_\bk^2}\;.
\ee
The authors of \cite{MarcoPRB2008} argued, in analogy with disordered $s$-wave superconductors, that the corrections brought by $J$ to the self-energy and to the vertex in the $\Delta_x$-channel of Eq.~\eqn{2} cancel each other also in the impurity model. If one then 
considers a model with Hamiltonian 
\be
\mathcal{H} = \mathcal{H}_\text{2AIM} - h_x\,\Delta_x\,,
\label{H2AIM-2}
\ee
see equations \eqn{H2AIM} and \eqn{2}, with a symmetry breaking term $h_x\ll J$, the impurity self-energy also becomes a two by two matrix 
with elements $\Sigma_{ab}(i\ep)$, with $a,b=1,2$ labelling the impurities. Following the above arguments one should expect that 
$h_x$ brings about a low energy scale $\Delta$ that cutoffs the 
singularities of $\Sigma(i\ep)$ in Eq.~\eqn{Sigma-1} so that
\be
\begin{split}
\Sigma_{11}(i\ep) &= i\ep - i\ep\,
Z\Big(i\,\sqrt{\ep^2+\Delta^2\,}\Big)^{-1}\,=\Sigma_{22}(i\ep)\,,\\
\Sigma_{12}(i\ep) &= \Delta\,
Z\Big(i\,\sqrt{\ep^2+\Delta^2\,}\Big)^{-1}\,=\Sigma_{21}(i\ep)\,.
\end{split}\label{Sigma-4}
\ee 
This ansatz was shown to fit extremely well the numerical data obtained in Ref.~\cite{LorenzoPRB2004} 
by directly solving the model in Eq.~\eqn{H2AIM-2} via the numerical renormalisation group. This result demonstrates, from a quite general perspective, how a pseudo-gapped symmetry invariant phase can nonetheless develop a sizeable symmetry breaking order parameter, which was used by the authors of Ref.~\cite{MarcoPRB2008} to interpret the phase diagram of a model that maps by DMFT onto the two-impurity model Eq.~\eqn{H2AIM}, which I describe below.

\subsection{The lattice model counterpart of the two-impurity model}  

Let us consider the two band Hubbard model in a infinitely coordinated Bethe lattice with Hamiltonian 
\be
\mathcal{H} = -\fract{t}{\sqrt{z}}\,\sum_{a=1}^2\,
\sum_{<i,j>\,\sigma} \Big(c^\dagger_{ai\sigma}c^\dagga_{aj\sigma}
+ H.c.\Big) + \fract{U}{2}\,\sum_i\,\big(n_i-2\big)^2
-2J\,\sum_{i}\,\bigg(T_{i\,x}^2 + T_{i\,y}^2\bigg)\,,
\label{H-exE}
\ee
where $z\to\infty$ is the coordination number, $n_i = n_{1i}+n_{2i}=\sum_{a=1}^2\sum_\sigma\,
c^\dagger_{ai\sigma}c^\dagga_{ai\sigma}$ is the charge density at site $i$ and $T_{i\,\alpha}$, $\alpha=x,y,z$, are the components 
of the orbital pseudo-spin $\mathbf{T}_i$ defined through 
\be
\mathbf{T}_i = \fract{1}{2}\,\sum_{a,b=1}^2\,\sum_\sigma\,
c^\dagger_{ai\sigma}\,\boldsymbol{\sigma}_{ab}\,
c^\dagger_{bi\sigma}\,,\label{T}
\ee
with $\boldsymbol{\sigma}$ the Pauli matrices. This model describes an $e\times E$ Jahn-Teller effect within the antiadiabatic approximation~\cite{exE-PRL2004}.\index{dynamical mean field theory!two-orbital $e\times E$ Jahn-Teller problem} Alternatively, one may rewrite 
Eq.~\eqn{H-exE} as 
\be
\begin{split}
\mathcal{H} &= -\fract{t}{\sqrt{z}}\,\sum_{a=1}^2\,
\sum_{<i,j>\,\sigma} \Big(c^\dagger_{ai\sigma}c^\dagga_{aj\sigma}
+ H.c.\Big) + \fract{U}{2}\,\sum_i\,\sum_{a=1}^2\,
\big(n_{ai}-1\big)^2\\
&+\sum_i\,\bigg(4J\,\mathbf{S}_{1i}\cdot\mathbf{S}_{2i} 
+ V\,n_{1i}\,n_{2i}\bigg)\,,
\end{split}\label{H-2HP}
\ee
where $V=U-J$, which represents two Hubbard models coupled by an antiferromagnetic exchange and by a charge repulsion.\\
Finally, if we interchange spin with orbital indices, the Hamiltonian 
\eqn{H-exE} can also be written as 
\be
\mathcal{H} = -\fract{t}{\sqrt{z}}\,\sum_{a=1}^2\,
\sum_{<i,j>\,\sigma} \Big(c^\dagger_{ai\sigma}c^\dagga_{aj\sigma}
+ H.c.\Big) + \fract{U}{2}\,\sum_i\,\big(n_i-2\big)^2
-2J\,\sum_{i}\,\bigg(S_{i\,x}^2 + S_{i\,y}^2\bigg)\,,
\label{H-SO}
\ee 
where now $S_{i\,\alpha}$ are the components of the total spin at site $i$, which is a two-band Hubbard model with a single-ion anisotropy that favours the spin to lie in the $x$-$y$ plane. \\

Let us for simplicity focus just on the Hamiltonian \eqn{H-exE}, or 
its equivalent representation \eqn{H-2HP}. The Hamiltonian contains three parameters, the conduction bandwidth $W=4t$, the Hubbard $U$ and the exchange $J$. The latter mediates pairing in the $s$-wave channel of Eq.~\eqn{3} that is however contrasted by $U$. The net  effect is a \textit{bare} scattering amplitude in that pairing channel $A_0=\rho_0\big(U-2J\big)$, where 
$\rho_0$ is the non-interacting DOS at the Fermi energy. At fixed $J\ll W$, as we shall assume hereafter, mean-field theory predicts   
upon increasing $U$ a BCS superconducting domain that extends from $U=0$, where pairing is maximum, to $U=2J$, where pairing disappears. Above $2J$ the ground state of the model \eqn{H-exE} should turn into  
a normal metal. This prediction is actually independent of the coordination number $z>2$ and the electron density.\\
When $z\to\infty$ 
this lattice model maps by DMFT onto the two-impurity model Eq.~\eqn{H2AIM} with the 
addition of a charge repulsion $V$ among the impurities. This repulsion is irrelevant and does not change the phase diagram of the impurity model, which thus remains similar to that in Fig.~\ref{phd}. 
If DMFT self-consistency is imposed, as $U$ increases at fixed 
$J\ll W$ and at half-filling, the lattice model is pushed towards a Mott transition, exactly like the single-band Hubbard model of section \ref{Ordinary Kondo physics at the Mott transition}. In the impurity language, this transition corresponds to $T_K\to 0$. However, still in the metal phase before that happens, $T_K$ will become comparable to $J$, even though the bare conduction bandwidth $W\gg J$. This is right the point where the impurity model crosses its QCP. Already before that happens, the impurity susceptibilities in the channels of equations \eqn{1}, \eqn{2} and \eqn{3} will be strongly enhanced and thus may drive, through the self-consistency, a true bulk instability. An instability in the first channel Eq.~\eqn{1} translates in the lattice model into an instability towards Ne\'el antiferromagnetic order, with the two orbitals occupied by opposite spin electrons. The instability in channel \eqn{2} corresponds instead to a cooperative Jahn-Teller effect. However both 
\eqn{1} and \eqn{2} are particle-hole channels and thus they require nesting of the Fermi surface to drive a bulk instability. 
On the contrary, the particle-particle channel in Eq.~\eqn{3} does not require any particular property of the Fermi surface to drive a superconducting instability, but just a finite DOS at the chemical potential. It is thus tempting to conclude that generically, i.e. in the absence of nesting and with finite DOS at Fermi, there must exist         
another superconducting dome besides the weak coupling 
$U<2J\ll W$ BCS one, right next to the Mott transition. 
\begin{figure}[t!]
 \centering
 \vspace{-1cm}
 \includegraphics[width=0.6\textwidth]{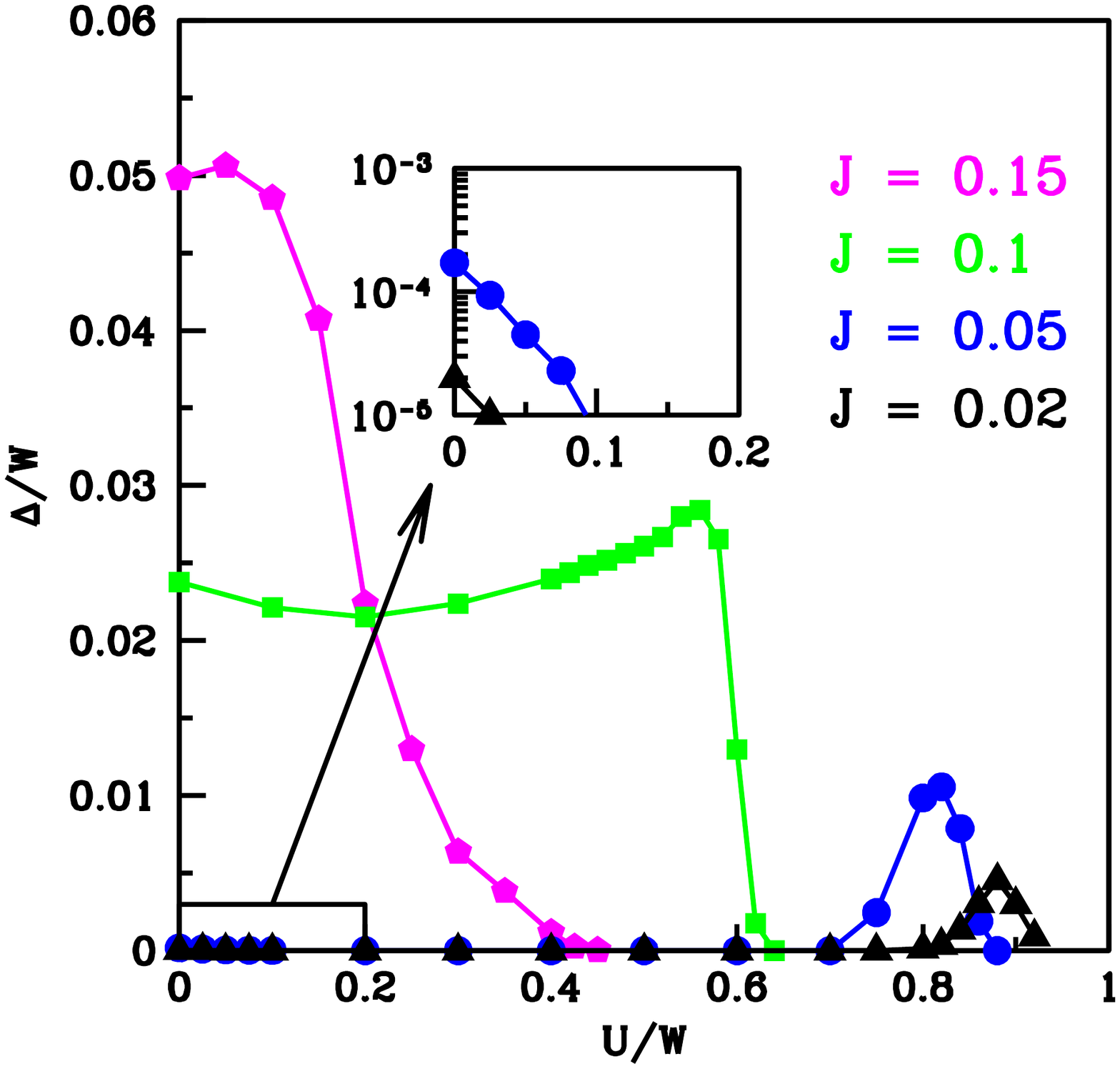}
 \vspace{-3cm}
 \caption{Superconducting gap of the model \eqn{H-exE} at electron density $n=2$ as function 
 of $U$ and for different values of $J$, in units of the conduction bandwidth 
 $W$. In the inset the weak coupling gap is zoomed. [From Ref.~\cite{exE-PRL2004}.]}
 \label{gap-exE}
\end{figure}
This expectation was confirmed by a full DMFT calculation in 
Ref.~\cite{exE-PRL2004}. In Fig.~\ref{gap-exE} the superconducting gap of the model \eqn{H-exE} is plotted at electron density $n=2$ as function of $U$ and for different values of $J$. I note that for 
the largest $J=0.15$, in units of the conduction bandwidth $W$, 
the gap is monotonically decreasing with $U$ and disappears at $U\simeq 2J$ where the superconductors turns by a weakly first order transition into a Mott insulator. This insulating phase is non-magnetic with all sites occupied by two electrons, one on each orbital, coupled into a spin-singlet configuration; a local version of a valence-bond crystal. Already for a weaker $J=0.1$ the gap becomes non monotonous; it first decreases than increases again before the first order Mott transition. For smaller $J=0.05$ and 
$J=0.02$, the superconducting phase splits, as anticipated, into two well separated domains. One appears at weak coupling and has a tiny BCS-like gap exponentially small in $1/J$, see the inset of Fig.~\ref{gap-exE}. However, another bell-shaped superconducting dome emerges at strong coupling next to the Mott transition and displays a huge gap if compared with the weak coupling BCS one; a striking example of superconductivity boosted by strong correlations ~\cite{SCS-Science2002}.\index{dynamical mean field theory!two-orbital $e\times E$ Jahn-Teller problem!strongly-correlated superconductivity}\\
The physics of the impurity model \eqn{H2AIM} allows anticipating 
not only the phase diagram but also the dynamical properties of the lattice model. Within DMFT one can prevent superconducting symmetry breaking and thus access the unstable zero-temperature normal phase, whose single-particle self-energy was 
found~\cite{exE-PRL2004} to be well fitted by that of the impurity model, see Eq.~\eqn{Sigma-1}.          
\begin{figure}[t!]
 \centering
 \vspace{-1cm}
 \includegraphics[width=0.6\textwidth]{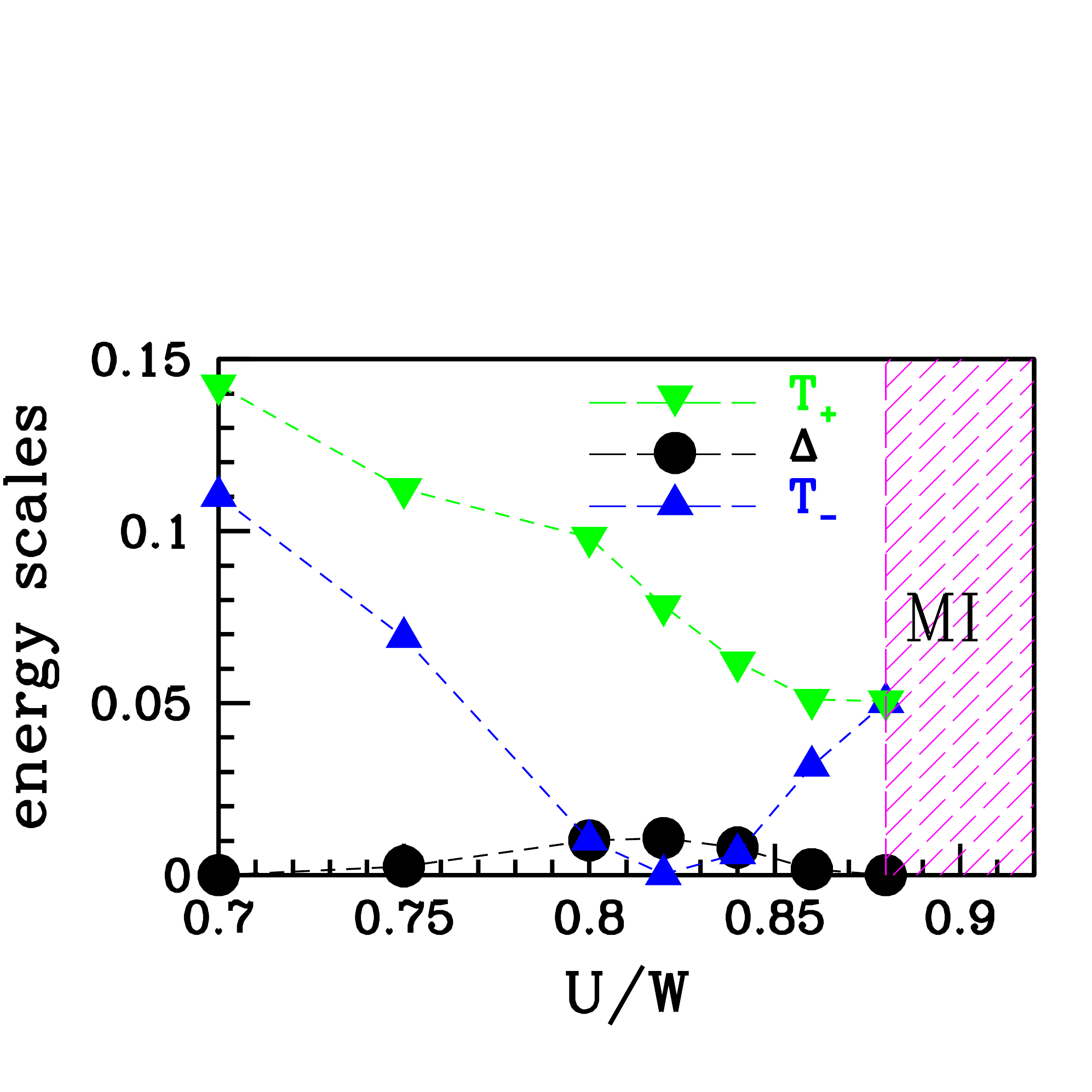}
 \caption{Impurity scales $T_+$ and $T_-$, see Eq.~\eqn{Sigma-1}, 
 extracted by the DMFT self-energy within the unstable normal phase, 
 in comparison with the superconducting gap that is obtained once 
 symmetry breaking is permitted. MI indicate the Mott insulating phase.
  [From Ref.~\cite{exE-PRL2004}.]}
 \label{scale}
\end{figure}
Fig.~\ref{scale} shows the impurity scales $T_+$ and $T_-$ extracted by fitting the DMFT self-energy within the unstable normal phase 
through Eq.~\eqn{Sigma-1}. We can observe that the impurity critical  
point is encountered before the Mott transition and, once one permits  
superconducting symmetry breaking, it corresponds to the maximum of the superconducting gap $\Delta$. I remark that the Mott insulator appears now when $T_+=T_-$ on the unscreened side of the impurity model, 
at which point the two Lorentzian that define the low-energy DOS
$\mathcal{N}_-(\ep)$, see Eq.~\eqn{N(ep)}, cancel each other. This is evidently different from the single-band case, where, as I mentioned, the transition occurs by the gradual disappearance of a Kondo-like resonance.\index{dynamical mean field theory!two-orbital $e\times E$ Jahn-Teller problem!connection with the two impurity model}

One can push the interpretation via the impurity model even further. 
Suppose we are in the Mott insulator at half-filling $n=2$, which, as mentioned, corresponds in the impurity model to the unscreened phase with $T_+=T_-$, and dope it, $n\to 2-\delta$. According to 
Eq.~\eqn{N(ep)-mu} we should expect that the insulator turns into a pseudo-gapped normal metal as soon as $\delta>0$. Moreover, upon increasing further the doping $\delta$, the impurity model should cross its quantum critical point, see Fig.~\ref{phd}, which should correspond in the lattice model after DMFT self-consistency to reappearance of a superconducting dome. Once again this expectation was indeed confirmed.       
\begin{figure}[t!]
 \centering
 \includegraphics[width=0.6\textwidth]{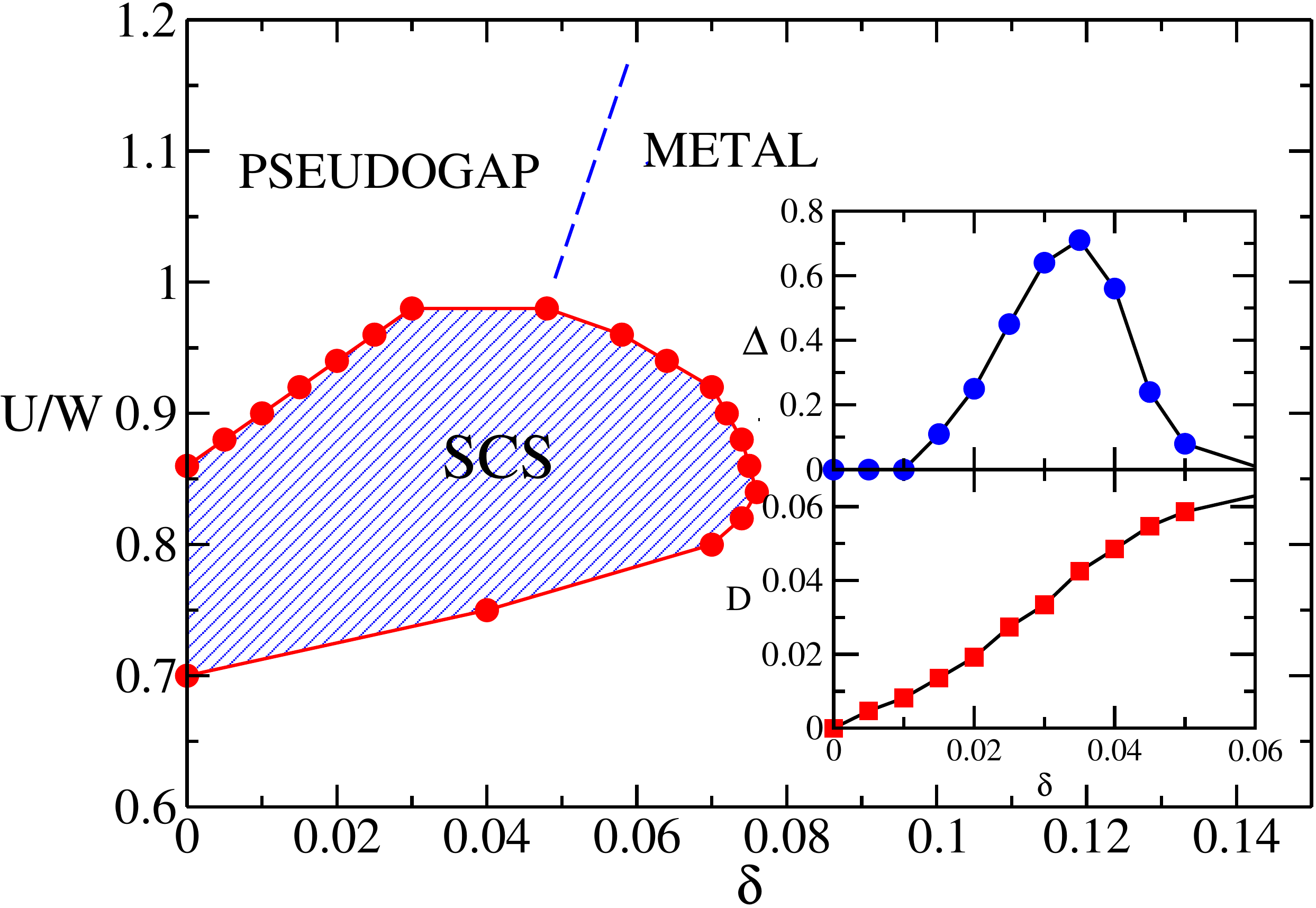}
 \caption{Phase diagram of the model Eq.~\eqn{H-exE} at $J=0.05$ as function of $U/W$ and doping $\delta$ away from half-filling. The insets show the dependence upon doping of the superconducting gap $\Delta$ and of the Drude weight $D$ at $U=0.92W$. 
  [From Ref.~\cite{exE-PRL2004}.]}
 \label{phd2}
\end{figure} 
Fig.~\ref{phd2} shows the final phase diagram at $J=0.05$ as function of $U/W$ and doping $\delta>0$. The superconducting dome extends at 
finite doping into a whole region such that, starting from the Mott insulator and doping it, one first finds a pseudo gapped normal phase that, upon further doping, turns into a superconductor that eventually disappears at higher doping into a well-behaved normal metal, i.e. not anymore pseudo gapped, see the insets of Fig.~\ref{phd2}.\index{dynamical mean field theory!two-orbital $e\times E$ Jahn-Teller problem!phase diagram}

\subsection{Landau-Fermi liquid picture}
It is impossible not to see striking similarities between the phase diagram in Fig.~\ref{phd2} and the phenomenology of high-$T_c$ copper oxides. This is even more evident by the behaviour of the Drude weight $D$, shown in the bottom inset of 
Fig.~\ref{phd2}, which grows linearly in $\delta$ upon doping the Mott insulator because of the linear-in-$\delta$ filling of the pseudo gap. \\
One may thus wonder whether and in which terms  
the predictive power of the impurity physics transferred to lattice models could be extended even beyond the limits of applicability of 
DMFT, namely even in realistic lattices with finite coordination number. I already mentioned how the emergence of a Kondo-like physics close to the Mott transition can be inferred quite generally from Landau-Fermi liquid theory. It is therefore worth addressing how one can translate the physics of the two impurity model in the language 
of Landau's Fermi liquids. Since a local Fermi liquid description can be defined also for impurity models~\cite{Nozieres1974}, it is convenient to start with that and eventually extend it to bulk systems.\index{Landau-Fermi liquid theory!local Fermi liquids} \\
Concerning the impurity model Eq.~\eqn{H2AIM}, both screened and unscreened phases are Fermi-liquid like. While this is evidently the case in the screened phase, where ordinary Kondo effect takes place, 
it is by far less obvious in the unscreened one that is characterised by a singular impurity self-energy. In Ref.~\cite{LorenzoPRB2004} it was shown that the conventional definition~\cite{Fred1978}
\[
\rho_\text{qp} =  \fract{\mathcal{N}(0)}{Z}\,,
\]
of the quasiparticle DOS at the impurity site and 
at the Fermi energy, where $\mathcal{N}(0)$ is the particle DOS at Fermi and $Z$ the quasiparticle residue, must be generalised to account for unscreened Fermi liquid phases into 
\index{Landau-Fermi liquid theory!local Fermi liquids!quasiparticle density of states}
\be
\rho_\text{qp} = \!\!\int \fract{d\ep}{\pi}\,
\fract{\partial f(\ep)}{\partial \ep}\, \text{Im}\Bigg\{
\mathcal{G}(\ep+i0^+)\,\Bigg[1-\left(\fract{\partial\Delta(\zeta)}{\partial\zeta}\right)_{\big|\zeta=\ep+i0^+}
\!\!-\left(\fract{\partial\Sigma(\zeta)}{\partial\zeta}\right)_{\big|\zeta=\ep+i0^+}\Bigg]\Bigg\}\,,
\label{rho_qp}
\ee
where
\[
\Delta(\zeta) = \int \fract{d\ep}{\pi}\, \fract{\Gamma(\ep)}{\zeta-\ep}
\simeq -i\,\Gamma(0)\,\text{sign}\left(\text{Im}\,\zeta\right)\;,
\]
is the Hilbert transform of the hybridisation function, $\mathcal{G}(\zeta)$ and $\Sigma(\zeta)$ the impurity Green's function and self-energy, respectively, continued in the complex frequency plane. In the screened phase of the model Eq.~\eqn{H2AIM} and for negligible $\partial \Delta(\zeta)/\partial\zeta$ one gets the 
conventional result
\[
\rho_\text{qp}{_{+}} = \fract{\mathcal{N}_+(0)}{Z} = \fract{1}{2\pi}\,\fract{T_+ +T_-}{T_+\,T_-}\;.
\]\\
In the pseudo gap phase, even though $Z$ vanishes, still $\rho_\text{qp}$ in Eq.~\eqn{rho_qp} has a well defined value since the singularity in the self-energy is cancelled by the vanishing DOS 
\[
\mathcal{N}_-(\ep) = -\fract{1}{\pi}\, \text{Im}\,\mathcal{G}(\ep+i0^+)\,,
\]
at $\ep=0$, leading to a finite \textit{quasiparticle} DOS at Fermi
\be
\rho_\text{qp}{_{-}} = \fract{1}{\pi}\,\fract{T_+ +T_-}{T_+\,T_-}\,,
\ee
despite the vanishing \textit{particle} DOS. This is remarkable, since common wisdom would rather suggest that a singular self-energy is incompatible with Landau's Fermi liquid theory.\\
In addition, even though a local Fermi liquid description does not apply right at the critical point $T_-=0$, still one can approach it from either Fermi liquid side of the transition. In particular, it was explicitly 
verified~\cite{LorenzoPRB2004} that the quasiparticle scattering amplitudes, 
defined in the screened phase through\index{Landau-Fermi liquid theory!local Fermi liquids!quasiparticle scattering amplitudes} 
\be
A_i \equiv Z^2\,\rho_\text{qp}\,\Gamma_i\,,\label{A-QIM}
\ee
where $\Gamma_i$ is the interaction vertex in channel $i$, tend to finite values approaching the critical 
point $T_-\to 0$, or equivalently $Z\to 0$, in all three relevant channels of equations~\eqn{1}--\eqn{3}, 
specifically $A\to 1$ in channels \eqn{1} and {2}, and $A\to -2$ in channel \eqn{3}. This confirms the expectation that vertex corrections cancel exactly self-energy ones, as assumed in the previously discussed Ref.~\cite{MarcoPRB2008}. \\

Suppose we could export the above local Fermi liquid results to the lattice model \eqn{H-exE}.  As I mentioned, the bare scattering amplitude in the $s$-wave Cooper channel Eq.~\eqn{3} is 
\[
A_0 = \rho_0\,\Big(U-2J\Big)\,,
\]
where $\rho_0$ is the non-interacting DOS at the Fermi energy. For $J\rho_0\ll 1$, as we assumed, 
and close to the Mott transition, the amplitude $A_0>0$ and thus one should not expect any superconductivity. In reality, the quasiparticle amplitude in that channel reads
\be
A = \rho_\text{qp}\,\Big(Z^2\,\Gamma_U - 2Z^2\,\Gamma_J\Big)\,.\label{A-qp}
\ee
The contribution from the charge channel $ Z^2\Gamma_U \sim ZU$ becomes negligible  
approaching the Mott transition; quasiparticles slow down and at the same time they undress from the strong repulsion. Moreover, just because they spend more time on each site before hopping to neighbouring ones, the quasiparticles can take more advantage of the local $J$-term, so that it is well conceivable 
that $ Z^2\Gamma_J\sim J$ is to a large extent unrenormalized by the proximity to a Mott transition, which once again entails cancellation of vertex and self-energy corrections. The outcome is that approaching the Mott transition  
\be
A\sim \rho_\text{qp}\,(ZU -2J)\simeq \rho_0\,\left(U-2\fract{J}{Z}\right)
\,,\label{A-simple}
\ee
changes sign from positive to negative, thus permitting a superconducting instability to set in despite the bare value 
$A_0$ does not, as indeed found by 
DMFT~\cite{exE-PRL2004,MarcoPRB2008}. Moreover, $A$ may now become of order $O(1)$ when 
$\rho_\text{qp}\,J\simeq \rho_0 J/Z\sim 1$, despite $\rho_0 J\ll 1$, suggesting that superconductivity is effectively pushed to the maximum $T_c\sim 0.055\,g$ attainable at a given pairing strength $g$,  again consistent with DMFT results. \\
In the impurity model the maximum of $A$ occurs right at the critical point, beyond which, in the unscreened phase, $A$ diminishes again~\cite{LorenzoPRB2004}. 
Here, as I mentioned, one should not use anymore Eq.~\eqn{A-QIM} to define the scattering amplitudes~\cite{LorenzoPRB2004}. This suggests that the simple expression Eq.~\eqn{A-simple} is only valid in the 
counterpart of the screened phase and cannot be pushed till the Mott transition, which would otherwise imply 
the unphysical result $A\sim 1/Z \to\infty$; some readjustment must intervene before, which in the impurity model is the pseudo gap opening that was also observed as a stable phase in the lattice model Eq.~\eqn{H-exE} away from half-filling. 

\section{Concluding remarks}

In conclusion, the Landau-Fermi liquid theory, in its original bulk formulation~\cite{Landau-1,Landau-2} as well as in its local version~\cite{Nozieres1974,Fred1978,LorenzoPRB2004}, seems to be 
the natural framework for building a bridge between the Kondo physics of impurity models and the Mott physics of lattice models, which connects to each other not just gross spectral features, like the Kondo resonance to the quasiparticle peak, but also more subtle properties of even greater impact, like the channels in which the impurity shows enhanced susceptibility to those in which the lattice model develops a true bulk instability. I have shown how this task can be explicitly accomplished in the case of the 
two-impurity model Eq.~\eqn{H2AIM} in connection with the lattice model Eq.~\eqn{H-exE} treated by DMFT, i.e. in the limit of infinite coordination number $z\to\infty$. The same approach has also been used to gain insights from the physics of a C$_{60}^{n-}$ impurity model~\cite{LorenzoPRL2005} 
into a model for alkali doped fullerides A$_3$C$_{60}$ that was studied by DMFT still in the $z\to\infty$ limit~\cite{FabrizioRMP2009},  but whose results reproduce quite well the physical properties of those molecular conductors. I further mention that the phase diagram of the lattice model Eq.~\eqn{H-exE} in one dimension, i.e. the opposite extreme to infinitely coordinated lattices, also recalls~\cite{FabrizioPRL2005} the same impurity phase diagram of Fig.~\ref{phd}, apart from some obvious differences. \\

It is therefore quite tempting to speculate that the relationship between impurity and lattice models close to a Mott transition remains even beyond the limit of infinitely coordinated lattices, with the due caution about  spatial correlations. As previously discussed, this connection can turn extremely fruitful if the impurity model upon increasing $U$ crosses a critical point, or gets very close to it,  i.e. it goes through a genuine phase transition or just a very sharp crossover. Any critical point is generically unstable in several symmetry-lowering channels, which share the property of being \textit{orthogonal} to charge that is instead severely suppressed by $U$. We could then argue that, in the corresponding lattice model upon approaching the Mott transition from the metallic side, a spontaneous symmetry breaking should intervene in one of the impurity instability channels. Which one dominates is going to be dictated by the spatial correlations that it entails 
with respect to the lattice structure and Hamiltonian parameters, besides its relevance relative to all other instability channels of the impurity. For instance, the instability in a particle-particle channel leading to superconductivity is less sensitive to the lattice structure than, e.g., a magnetic instability. However the coupling constant in the magnetic channel is inevitably stronger than that in the pairing channel, especially as the system gets closer to the Mott transition. The outcome of such competing effects might be that superconductivity may appear first and then gives way to magnetism, sooner or later depending on the degree of magnetic frustration, as shown e.g. for doped fullerides~\cite{FabrizioRMP2009}, or it may be defeated by magnetism and not appear at all. Even in that case, superconductivity can re-emerge upon doping the magnetic insulator. \\
There is actually a plethora of impurity models whose rich phase diagrams could translate into equally rich phase diagrams of corresponding lattice models. The issue is whether those lattice models are realistic and can describe physical systems. For instance, 
clusters of Anderson impurities~\cite{Nutshell} could be used to interpret the results of cluster DMFT calculations~\cite{MaltePRB2016}, even though the $N$-site extension of DMFT  
is only exact for $N\to\infty$ and therefore finite $N$ calculations could be biased by the small cluster size. 



\clearpage

\bibliographystyle{correl}

\clearchapter


\end{document}